\documentclass[apj]{emulateapj}
\usepackage{apjfonts}
\usepackage{graphicx}
\usepackage{amssymb}
\usepackage{amsbsy}
\usepackage{amsmath}
\usepackage{mathrsfs}
\usepackage{color}
\usepackage[colorlinks, allcolors=black]{hyperref}

\DeclareMathAlphabet{\mathbi}{OT1}{ptm}{bx}{it}
\SetMathAlphabet\mathbi{bold}{OT1}{ptm}{bx}{it}

\slugcomment{{\it The Astrophysical Journal}, 831:206, 2016, November 10}
\lefthead{\sc{Li et al.}}
\righthead{\sc{A Non-parametric Approach to Constrain the TF in RM}}

\begin{document}
\title{A Non-parametric Approach to Constrain the Transfer Function in Reverberation Mapping}
\author{
  Yan-Rong Li\altaffilmark{1}, Jian-Min Wang\altaffilmark{1,2}, and Jin-Ming Bai\altaffilmark{3,4}
  }
\affil
{
$^1$ Key Laboratory for Particle Astrophysics, Institute of High 
Energy Physics, Chinese Academy of Sciences, 19B Yuquan Road, 
Beijing 100049, China; \href{mailto:liyanrong@mail.ihep.ac.cn}{liyanrong@mail.ihep.ac.cn}\\
$^2$ National Astronomical Observatories of China, Chinese 
Academy of Sciences, 20A Datun Road, Beijing 100020, China\\
$^3$ Yunnan Observatories, Chinese Academy of Sciences, Kunming 650011, China\\
$^4$ Key Laboratory for the Structure and Evolution of Celestial Objects, Chinese Academy of Sciences, 
Kunming 650011, China
}

\begin{abstract}
Broad emission lines of active galactic nuclei stem from a spatially extended
region (broad-line region, BLR) that is composed of discrete 
clouds and photoionized by the central ionizing continuum. The temporal behaviors 
of these emission lines are blurred echoes of the continuum variations 
(i.e., reverberation mapping, RM) and directly reflect the structures 
and kinematic information of BLRs through the so-called transfer function 
(also known as the velocity-delay map). 
Based on the previous works of \citeauthor{Rybicki1992} and \citeauthor{Zu2011}, 
we develop an extended, non-parametric approach to determine 
the transfer function for RM data, in which
the transfer function is expressed as a sum of a family of relatively displaced 
Gaussian response functions. Therefore, arbitrary shapes of transfer functions associated with 
complicated BLR geometry can be seamlessly included, enabling us to relax the 
presumption of a specified transfer function frequently
adopted in previous studies and to let it be determined by observation data. 
We formulate our approach in a previously well-established framework that 
incorporates the statistical modeling of the continuum variations as a damped random
walk process and takes into account the long-term secular variations which are irrelevant 
to RM signals. The Application to RM data shows the fidelity of our approach.
\end{abstract}

\section{Introduction}
The broad-line regions (BLRs) in active galactic nuclei (AGNs), with sizes of light days to 
light weeks (\citealt{Kaspi2000, Bentz2013}), are too compact to be spatially resolved with 
the existing telescopes, except in a few cases in which the BLRs are magnified by gravitational 
lensing (\citealt{Sluse2012, Guerras2013}). The reverberation mapping (RM) technique provides 
a powerful and unique way to probe the structures and kinematics of BLRs through analyzing the 
temporal variation patterns of emission lines and the ionizing continuum (\citealt{Blandford1982, 
Peterson1993}). Over the past four decades, RM studies have successfully measured BLR sizes for 
$\sim60$ nearby Seyfert galaxies and quasars (e.g., \citealt{Bentz2013, Du2015}). 
Particularly influential is the discovery that BLR sizes closely correlate with optical 
luminosities of (sub-Eddington) AGNs, as expected from simple photoionization theory 
(e.g., \citealt{Kaspi2000, Kaspi2005, Bentz2013, Du2015}).
This relationship, in combination with the velocity widths of broad emission lines,
opens up an economic way for the virial mass estimation of supermassive black holes (e.g., 
\citealt{Vestergaard2006, Ho2015}). Meanwhile, velocity-binned RM analysis for several AGNs 
with high-quality spectroscopic data has crudely seen signatures for diverse kinematics in BLRs, 
including radial inflows/outflows and non-radial, virialized Keplerian rotations 
(\citealt{Bentz2009, Denney2010, Grier2013b, Peterson2014, Du2016}). 

Despite these advances, the detailed information for structures and kinematics of BLRs 
remain largely unknown. 
The major challenges are from RM data sets reliably inferring the transfer function or 
velocity-delay map and from obtaining the structure and kinematic information of BLRs from this map. 
Traditional cross-correlation analysis (e.g., \citealt{Bentz2013}) or the previous 
method using a simple specified transfer function (e.g., \citealt{Rybicki1994, Zu2011, Zu2016})%
\footnote{\cite{Zu2011} presumed a top-hat transfer function and developed a software
called \texttt{JAVELIN}, available at http://www.astronomy.ohio-state.edu/{\textasciitilde}yingzu/codes.html\#javelin.}
yields no more than the characterized BLR sizes.

\begin{figure*}[!t]
\centering
\includegraphics[angle=-90.0, width=0.7\textwidth]{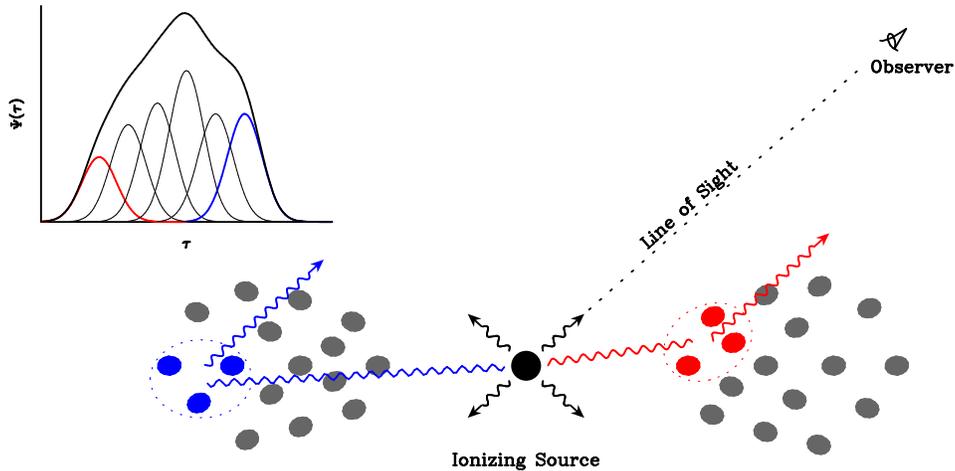}
\caption{Schematic of the transfer function for a system that consists of discrete clouds. The 
transfer function can be expressed as a sum of many relatively displaced response functions of
local adjacent clouds. For illustrative purposes, the red (blue) curve in the transfer
function corresponds to the response from the clouds highlighted in red (blue).}
\label{fig_sch_loc}
\end{figure*}

\begin{figure*}[!t]
\centering
\includegraphics[width=0.8\textwidth]{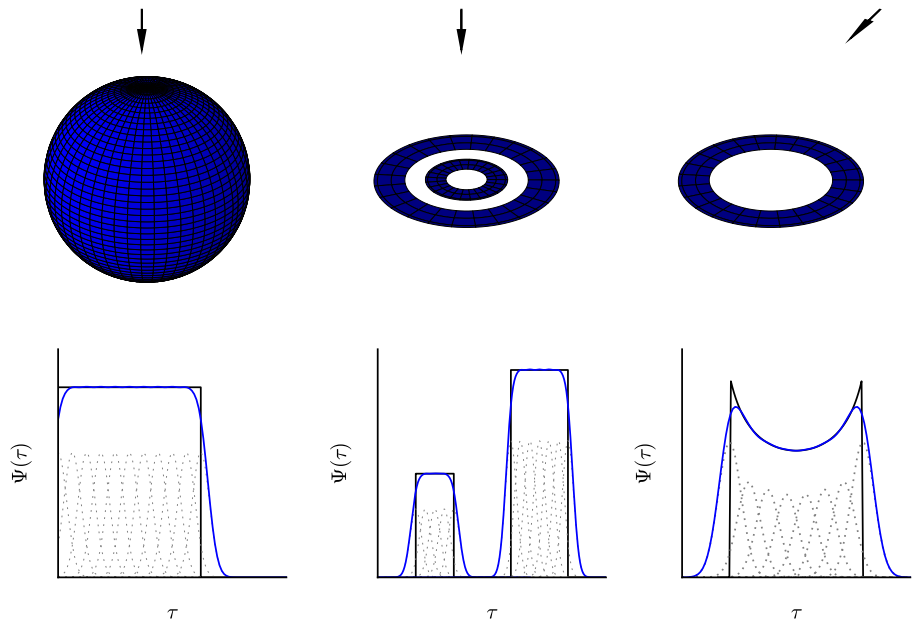}
\caption{Illustrative examples for three types of BLR geometry in which the transfer functions are
approximated by a family of Gaussian response functions: ({\it left}) a spherical shell, 
({\it middle}) two disk rings,  and ({\it right}) an inclined disk ring. Arrows indicate the line of sight.
In the bottom panels, black lines represent
the realistic transfer functions, and blue lines represent the sum of the Gaussian functions (gray
dotted lines). Our approach can particularly cope with the cases of 
multimodal or asymmetric transfer functions (e.g., middle panels), in which the previous 
\texttt{JAVELIN} method may encounter difficulties (see Appendix C).}
\label{fig_sch}
\end{figure*}

Great effort has gone into developing more sophisticated methods. There are generally two 
classes of methods depending on the strategy adopted in solving the integral equation 
that links the variations of continuum and emission lines in RM. 
One class relies on directly inverting the integral equation, including early attempts based 
on the Fourier transform (\citealt{Blandford1982}) and regularized linear inversion (RLI) methods 
(\citealt{Krolik1995, Skielboe2015}). 
The method using the Fourier transform was found to 
require high-fidelity data and thus has not seen much application (\citealt{Maoz1991}), 
whereas RLI seeks to discretize the integral equation and employ a differencing operator 
to suppress the noise. The downside of this class of methods 
is that the measurement noises cannot be self-consistently incorporated, 
thus, this class may be very noise sensitive in some cases. The other class relies on
indirectly inferring the ``best'' transfer function that fits the data 
by supposing prior limits or models on it. The maximum entropy method 
(\citealt{Horne1994}) and BLR dynamical modeling method (\citealt{Pancoast2011, Pancoast2014a, Li2013})
belong to this class.
The maximum entropy method finds the ``simplest'' solutions for the transfer function by maximizing the 
entropy on the premise of a reasonable fit to the data. However, it does not allow for
straightforward uncertainty estimates or model selection (\citealt{Horne2003}). The BLR dynamical modeling 
method presumes a specific but sufficiently flexible toy model for BLRs and directly infers
the model parameters; therefore, the transfer function is only an intermediate product
(\citealt{Pancoast2011, Li2013}).
It cannot yet satisfactorily assess the systematic errors of the adopted model.

In this paper, we describe an extended, non-parametric method to determine the transfer 
function from RM data, based on the previous works of \cite{Rybicki1992} and \cite{Zu2011}. 
Motivated by the scenario that BLRs are composed of a large number of clouds so that 
a local adjacent collection of these clouds effectively respond to the central ionizing continuum
as a Gaussian response function, we express the transfer function as a sum of a family of 
relatively displaced Gaussian functions. Therefore, we can account for arbitrary shapes of the transfer
function, thereby largely relaxing the need for presuming the transfer function (e.g., 
\citealt{Rybicki1994, Zu2011}) or
BLR models (e.g., \citealt{Pancoast2011, Li2013}). In this respect, our method is model 
independent and is particularly able to cope with complicated shapes of transfer functions (e.g., 
multimodal) for diverse BLR structures and kinematics.

The paper is organized as follows. We describe the methodology of our approach in
detail in Section 2 and perform simulation tests in Section~3.   Section~4
presents application of our approach to three RM AGNs to illustrate its fidelity. 
A discussion and conclusions are given in Section~5.
All the time delays quoted in the text are given in the observer's frame.

\section{Methodology}
\subsection{The Transfer Function}
In RM, variations of broad emission lines $f_l(t)$ are blurred echoes of
continuum variations $f_c(t)$ through the transfer function $\Psi(\tau)$ as
(\citealt{Blandford1982, Peterson1993})
\begin{equation}
f_l (t) = \int_{-\infty}^\infty \Psi(\tau) f_{c}(t-\tau) d\tau,
\label{eqn_line}
\end{equation}
where the integral limits are set to account for all possible positively and
negatively delayed responses (e.g, \citealt{Marshall2008,Shappee2014}).
Here, we show only velocity-unresolved RM so that velocity information 
is not included in the transfer function. However, it is easy to extend 
our approach to velocity-resolved RM by applying 
Equation~(\ref{eqn_line}) to different velocity bins of the emission lines
(see below for a discussion).
In addition, we also neglect the nonlinearity of the response for simplicity 
(\citealt{Li2013}). Nevertheless, we will provide below a way to incorporate non-linear 
response.  

Mathematically speaking, it is known that an arbitrary function can be expressed as a sum of 
many slightly displaced $\delta-$functions---namely, $f(x)=\int f(a)\delta(x-a) d a$, where 
$\delta(x)$ is the Dirac delta function. In practical calculations, one instead uses a set of 
basic functions with spatial extent to 
describe a smooth function. A general adoption is Gaussian functions, which are often found to be
adequate (\citealt{Sivia2006}, p.140). In this spirit, the transfer function in RM is now written as 
\begin{equation}
 \Psi(\tau)= \sum_{k=1}^{K} f_k \exp\left[-\frac{(\tau-\tau_k)^2}{2\omega^2}\right],
 \label{eqn_tf}
\end{equation}
where $\omega$ represents the common width of Gaussian responses, $\tau_k$ and $f_k$ are 
the mean lag and weight of $k$th response, and $K$ is the number of responses, 
which can be regarded as a ``smoothing parameter.'' The Gaussian functions
are fixed to a common width for the sake of simplicity. Such a choice also
helps to relax the parameter degeneracy and improve the computational efficiency of the
Bayesian inference described below. 

The physical interpretation for such an expression is straightforward. The system (i.e., BLRs) 
under consideration is composed of a large number of clouds. Each cloud responds to the variation 
of the central source as a $\delta-$response function. A collection of these local adjacent clouds 
effectively produce a Gaussian response function with a width. This is similar to the principle of the 
local optimally emitting cloud model, widely applied in photon-ionization calculations 
(\citealt{Baldwin1995}). Figure~\ref{fig_sch_loc} shows a schematic illustration of the 
transfer function of BLRs. In Figure~\ref{fig_sch}, we show examples for three types of BLR geometry---a spherical shell, double disk rings, and an inclined disk ring---in which the transfer function is approximated by
a family of Gaussian functions. 
In particular, there is evidence that H$\beta$ BLRs have two distinct components
(e.g., \citealt{Brotherton1996,Sulentic2000,Hu2008,Hu2012} and references therein), 
which indicate that their transfer functions may be bimodal, as illustrated 
in the middle panels of Figure~\ref{fig_sch}.

\subsection{Bayesian Inference}

With Equation~(\ref{eqn_tf}), we can directly
use the framework developed by \cite{Rybicki1992} and \cite{Zu2011} to infer 
the best transfer function. We stress that all the formulae presented in this 
section have been well established by \cite{Zu2011}. We list them below for 
the sake of completeness.

Continuum variation can be well described by a damped random walk (DRW) process
(e.g., \citealt{Kelly2009, Kelly2014, Zu2011, Li2013}).
A DRW process is a stationary process such that its covariance function at any two times 
$t_i$ and $t_j$ only depend on the time difference $\Delta t =t_i-t_j$, which can 
be simply prescribed as  
\begin{equation}
 S_{cc}(\Delta t) = \langle f_c(t_i)f_{c}(t_j)\rangle = 
 \sigma_{\rm d}^2\exp\left(-\frac{|\Delta t|}{\tau_{\rm d}}\right),
\label{eqn_cov_cc}
\end{equation}
where angle brackets represent the statistical ensemble average, $f_c(t)$ is 
the continuum driven by the DRW process, 
$\tau_{\rm d}$ represents the typical damping timescale, and  $\sigma_{\rm d}$ 
represents the standard deviation of the process on a long timescale ($\gg\tau_{\rm d}$). On a short
timescale, the variation amplitude of the process is $\sigma_{\rm d}\sqrt{2t/\tau_{\rm d}}$.
If this is combined with Equation~(\ref{eqn_line}), the covariance between the line and the continuum can be
written as
\begin{equation}
S_{lc}(\Delta t)=\langle f_l(t_i)f_{c}(t_j)\rangle = 
\int_{-\infty}^\infty \Psi(\tau) S_{cc}(\Delta t-\tau) d\tau,
\label{eqn_cov_lc}
\end{equation}
and the covariance of the line is 
\begin{eqnarray}
S_{ll}(\Delta t)&=&\langle f_l(t_i)f_{l}(t_j)\rangle\nonumber\\
&=& \int_{-\infty}^\infty\int_{-\infty}^\infty \Psi(\tau) \Psi(\tau') 
S_{cc}[\Delta t -(\tau-\tau')] d\tau d\tau'.
\label{eqn_cov_ll}
\end{eqnarray}
In Appendices A and B, we present a detailed derivation for the analytical expressions of these covariances
in terms of the error function.

The Bayesian posterior probability is obtained as follows (see \citealt{Rybicki1992} 
and \citealt{Zu2011} for a thorough derivation). Let $\mathbi{y}$ be a column vector 
comprised of the light curves of both the continuum and the line.  
A set of measurements in a campaign is equal to the realization of some underlying signals 
with measurement noises as
\begin{equation}
\mathbi{y} = \mathbi{s} + \mathbi{Lq} + \mathbi{n},
\end{equation}
where $\mathbi{s}$ is the signal of the variations, of which the portion for the continuum is described by
a DRW process, and $\mathbi{n}$ represents the measurement noises. Here the term $\mathbi{Lq}$ is usually
used to model general trends linearly varying with time in the light curves (\citealt{Rybicki1992}),
where $\mathbi{L}$ is a matrix of known coefficients and $\mathbi{q}$ is a column vector of 
unknown linear fitting parameters. For example, to remove separate means of the light
curves with a total of $N+M$ measurements, one just configures $\mathbi{L}$ to 
be a $2\times (N+M)$ matrix, where $N$ and $M$ are the number of measurements for 
the continuum and emission line, respectively. The matrix $\mathbi{L}$ has entries of $(1, 0)$
for the continuum data points and $(0, 1)$ for the line data points 
(\citealt{Rybicki1992, Zu2013}). If one desires to detrend the light curves with long-term 
secular variations using one-order polynomials, one needs to configure $\mathbi{L}$
as a $4\times (N+M)$ matrix (see below for details). Unless stated otherwise,
we remove the means of the light curves by default in the following calculations.

The signal $\mathbi{s}$ and noise $\mathbi{n}$ are assumed to be
Gaussian and independent, so that their probabilities are simply written as
\begin{equation}
P(\mathbi{s}) \propto \frac{1}{\sqrt{|\mathbi{S}|}}
\exp\left(-\frac{\mathbi{s}^T\mathbi{S}^{-1}\mathbi{s}}{2}\right),
\end{equation}
and 
\begin{equation}
P(\mathbi{n})\propto \frac{1}{\sqrt{|\mathbi{N}|}}
\exp\left(-\frac{\mathbi{n}^T\mathbi{N}^{-1}\mathbi{n}}{2}\right),
\end{equation}
respectively, where $\mathbi{S}=\langle\mathbi{s}\mathbi{s}\rangle$ is the covariance 
matrix of $\mathbi{s}$ given by Equations~(\ref{eqn_cov_cc})-(\ref{eqn_cov_ll}) and 
$\mathbi{N}=\langle\mathbi{nn}\rangle$ is the covariance matrix 
of $\mathbi{n}$. There is no prior information for the linear fitting parameters 
$\mathbi{q}$, and we assign the vector a uniform prior probability. Now the probability for 
a realization $\mathbi{y}$ is 
\begin{equation}
P(\mathbi{y|\sigma, \tau_{\rm d}},\boldsymbol{\theta}) 
\propto \iiint P(\mathbi{s})P(\mathbi{n})\delta[\mathbi{y-(s+n+Lq)}]
d\mathbi{s}d\mathbi{n}d\mathbi{q},
\end{equation}
where $\boldsymbol{\theta}$ represents the parameter set for the transfer function in Equation~(\ref{eqn_tf})
and the probability $P(\mathbi{q})$ is uniform, so that it is neglected in the above integral.
With some mathematical implementation, it can be shown that (\citealt{Zu2011}) 
\begin{equation}
P(\mathbi{y|\sigma, \tau_{\rm d}},\boldsymbol{\theta}) = 
\frac{1}{\sqrt{|\mathbi{C}||\mathbi{L}^T\mathbi{C}^{-1}\mathbi{L}|}}
\exp\left(-\frac{\mathbi{y}^T\mathbi{C}^{-1}_\perp\mathbi{y}}{2}\right),
\label{eqn_likeli}
\end{equation}
where $\mathbi{C=S+N}$ and $\mathbi{C}^{-1}_\perp=
\mathbi{C}^{-1}-\mathbi{C}^{-1}\mathbi{L}(\mathbi{L}^T\mathbi{C}^{-1}\mathbi{L})^{-1}\mathbi{L}^T\mathbi{C}^{-1}$.
The best estimate for $\mathbi{s}$ is (\citealt{Rybicki1992})
\begin{equation}
\mathbi{\hat s} = \mathbi{SC}^{-1}(\mathbi{y-L\hat q}),
\label{eqn_s}
\end{equation}
and the best estimate for the linear fitting parameter is
\begin{equation}
\mathbi{\hat q} = (\mathbi{L}^T\mathbi{C}^{-1}\mathbi{L})^{-1}\mathbi{L}^T\mathbi{C}^{-1}\mathbi{y}.
\label{eqn_q}
\end{equation}
The corresponding variance of the best estimate is, respectively, 
\begin{equation}
\langle\Delta \mathbi{s}^2\rangle = \mathbi{S} - \mathbi{SC}_{\perp}^{-1}\mathbi{S},
\end{equation}
and 
\begin{equation}
\langle\Delta \mathbi{q}^2\rangle =  (\mathbi{L}^T\mathbi{C}^{-1}\mathbi{L})^{-1}.
\end{equation}
The best estimate for the light curves is 
\begin{equation}
\mathbi{\hat y} = \mathbi{\hat s} + \mathbi{L\hat q},
\label{eqn_y}
\end{equation}
and the corresponding variance is
\begin{eqnarray}
\langle\Delta \mathbi{y}^2\rangle &=& \mathbi{S} - \mathbi{SC}^{-1}\mathbi{S}\nonumber\\
&&+(\mathbi{SC}^{-1}\mathbi{L} - \mathbi{L}) (\mathbi{L}^T\mathbi{C}^{-1}\mathbi{L})^{-1}
(\mathbi{SC}^{-1}\mathbi{L} - \mathbi{L})^{T}.
\end{eqnarray}
Note that Equations~(\ref{eqn_s}), (\ref{eqn_q}), and (\ref{eqn_y}) also provide a way to estimate the value 
of the light curves at some specified time that may not be one of the already measured points.  

According to the Bayes's theorem, the posterior probability distribution for 
the free parameter set is given by
\begin{equation}
P(\sigma, \tau_{\rm d},\boldsymbol{\theta}|\mathbi{y}) =\frac{ P(\sigma, \tau_{\rm d}, \boldsymbol{\theta}) 
P(\mathbi{y}|\sigma, \tau_{\rm d},\boldsymbol{\theta})}{P(\mathbi{y})},
\label{eqn_post}
\end{equation}
where $P(\mathbi{y})$ is the marginal likelihood that serves as a normalization factor to 
the posterior probability and $P(\sigma, \tau_{\rm d}, \boldsymbol{\theta})$ is the prior probability
for the parameters.
We maximize this posterior distribution to obtain the best estimate for the free parameters
using a Markov-chain Monte Carlo (MCMC) method as described in the next section.

Long-term secular variability is incidentally detected over the duration of the campaign in some 
RM objects (e.g., \citealt{Denney2010, Li2013, Peterson2014b}). 
Such long-term trends
are uncorrelated with reverberation variations and may bias the reverberation analysis
(\citealt{Welsh1999}).
The linear fitting parameter $\mathbi{q}$ provides a natural way to remove such trends.
If the long-term trends vary linearly with time ($q_1 + q_2 t$), the matrix $\mathbi{L}$ has the form of
(\citealt{Rybicki1992, Zu2011})
\begin{equation}
 \mathbi{L} =\left[
 \begin{array}{cccc}
  1 & t_1^c & 0 & 0\\
  \vdots & \vdots &\vdots & \vdots \\
  0 & t_N^c & 0 & 0\\
  0 & 0 & 1 & t_1^l\\
  \vdots & \vdots &\vdots & \vdots \\
  0 & 0 & 1 & t_M^l\\
 \end{array}
 \right],
\end{equation}
where $t_i^c$ is the $i$th time of the $N$ measurement points of the continuum and 
$t_i^l$ is the $i$th time of the $M$ measurement points of the line. In this case, $\mathbi{q}$ is a 
column vector with four entries, which represent
the coefficients of the linear trends for the continuum and line data.
For the long-term trends with high-order 
polynomial variation, one can similarly write out the matrix $\mathbi{L}$
by just adding extra columns. In our calculations, when the light curves
of the emission line and the driven continuum visually show different long-term trends, 
we switch on the detrending procedure by assuming the trends vary linearly with time
(see below for application to the RM data of Mrk 817).

\begin{figure*}[th!]
\centering
\includegraphics[width=0.8\textwidth]{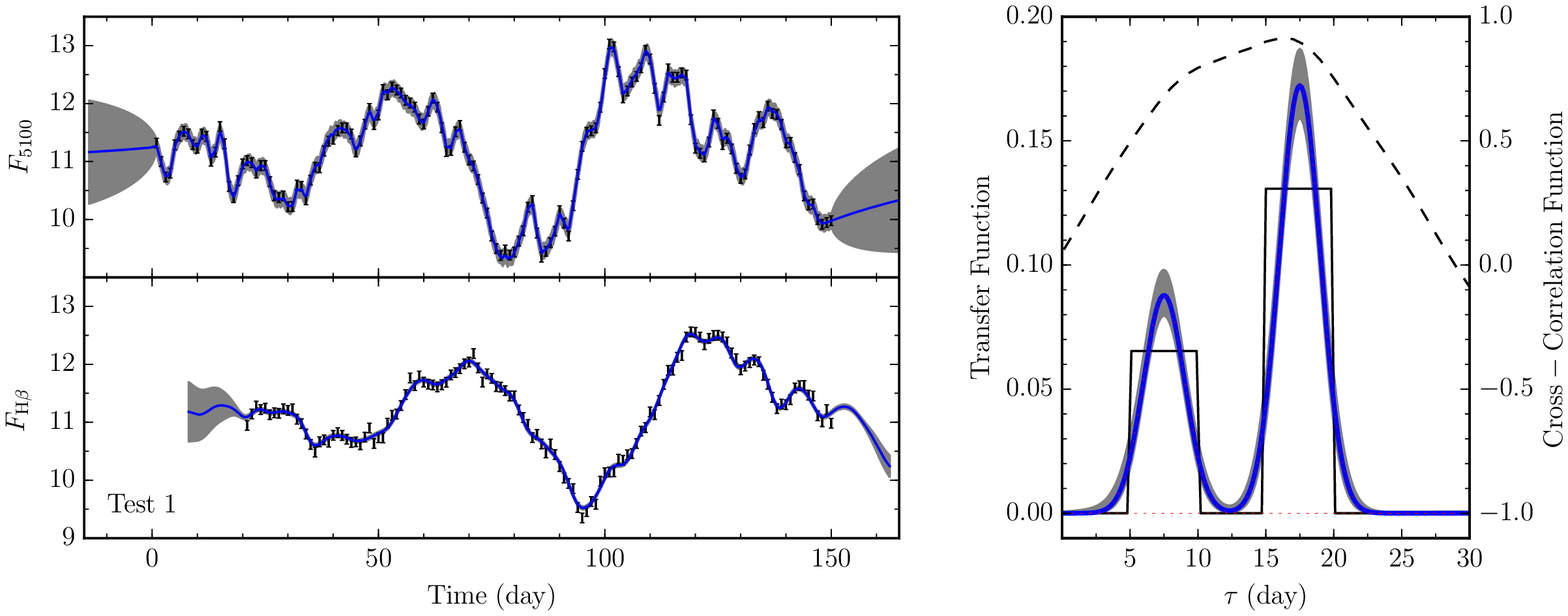}
\includegraphics[width=0.8\textwidth]{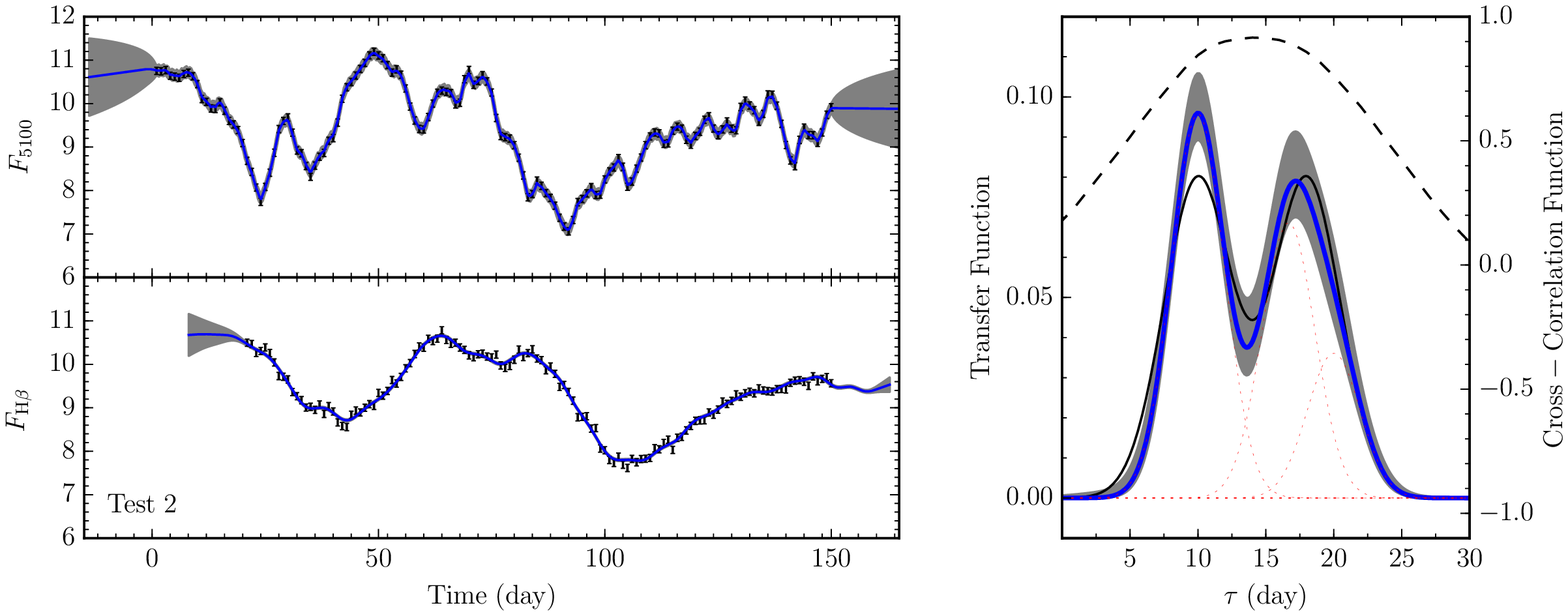}
\includegraphics[width=0.8\textwidth]{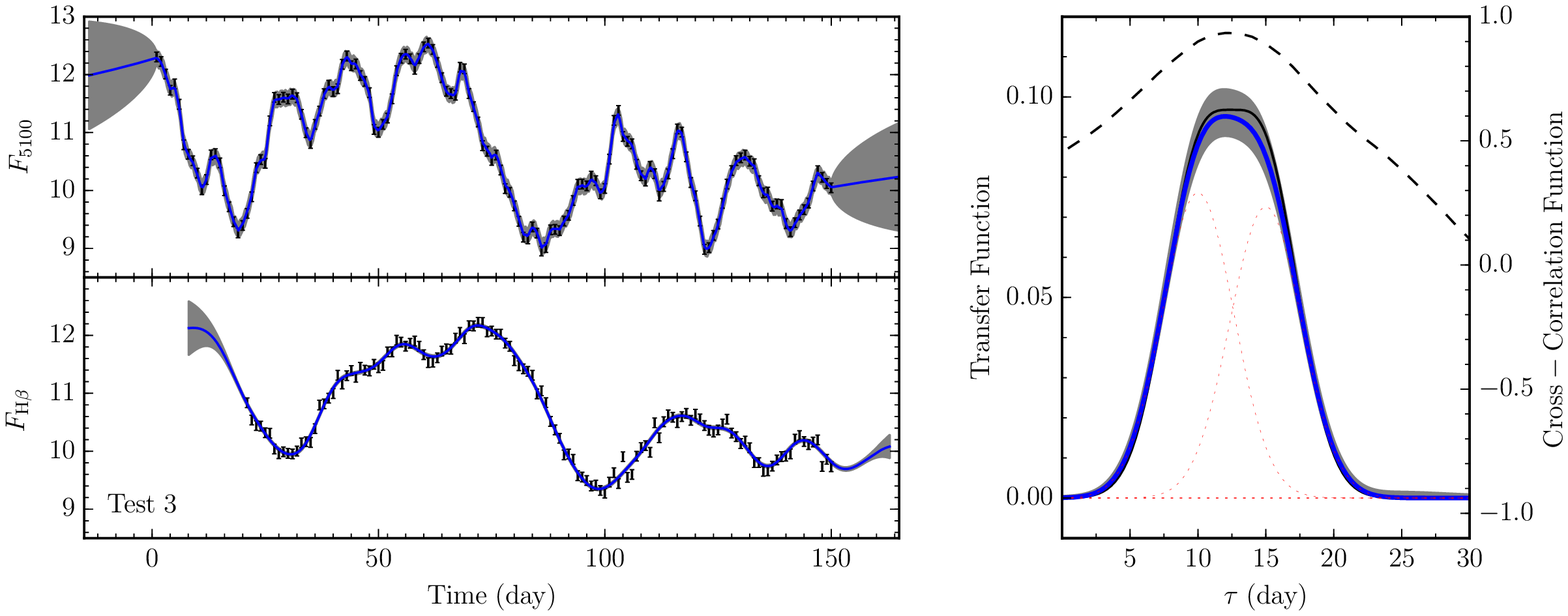}
\caption{Three simulation tests of our approach. Left panels: simulated light 
curves for 5100~{\AA} continuum and H$\beta$ emission line. Solid lines represent the recovered light curves, 
and shaded areas represent their uncertainties.
Right panel: the best recovered transfer function (in blue) compared with
the input transfer function (in black). 
The shaded area represents the estimated uncertainties for the 
recovered transfer function. Red dotted lines represent the Gaussian functions that constitute 
the transfer function. Dashed lines represent the interpolation cross-correlation function of the 
light curves.
}
\label{fig_sim}
\end{figure*}

\begin{figure*}[th!]
\centering
\includegraphics[width=0.3\textwidth]{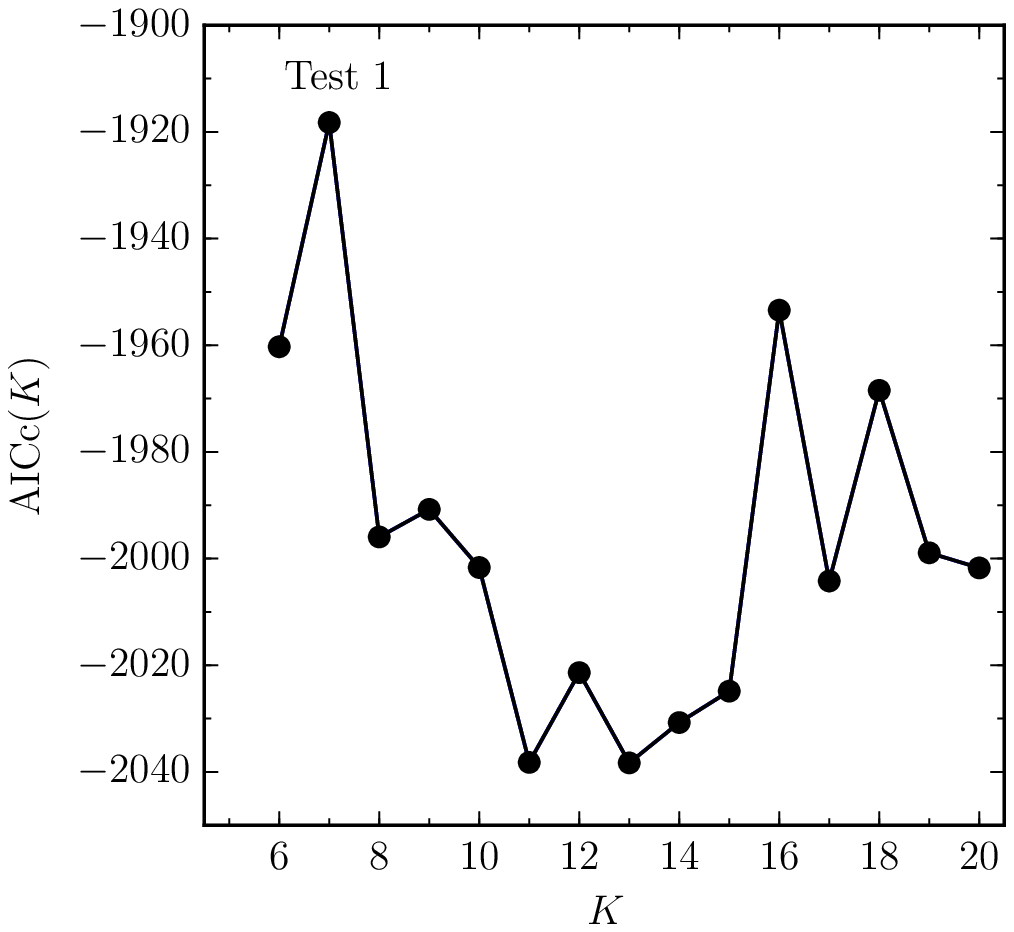}
\includegraphics[width=0.3\textwidth]{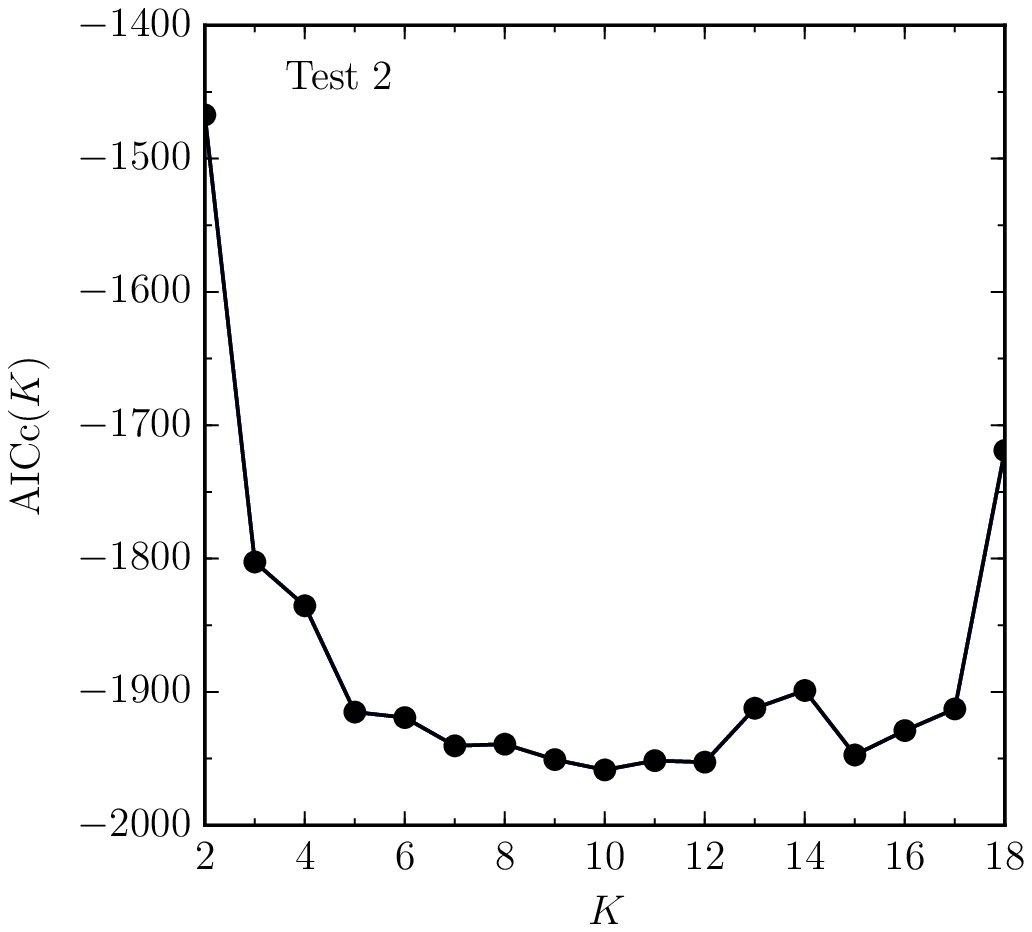}
\includegraphics[width=0.3\textwidth]{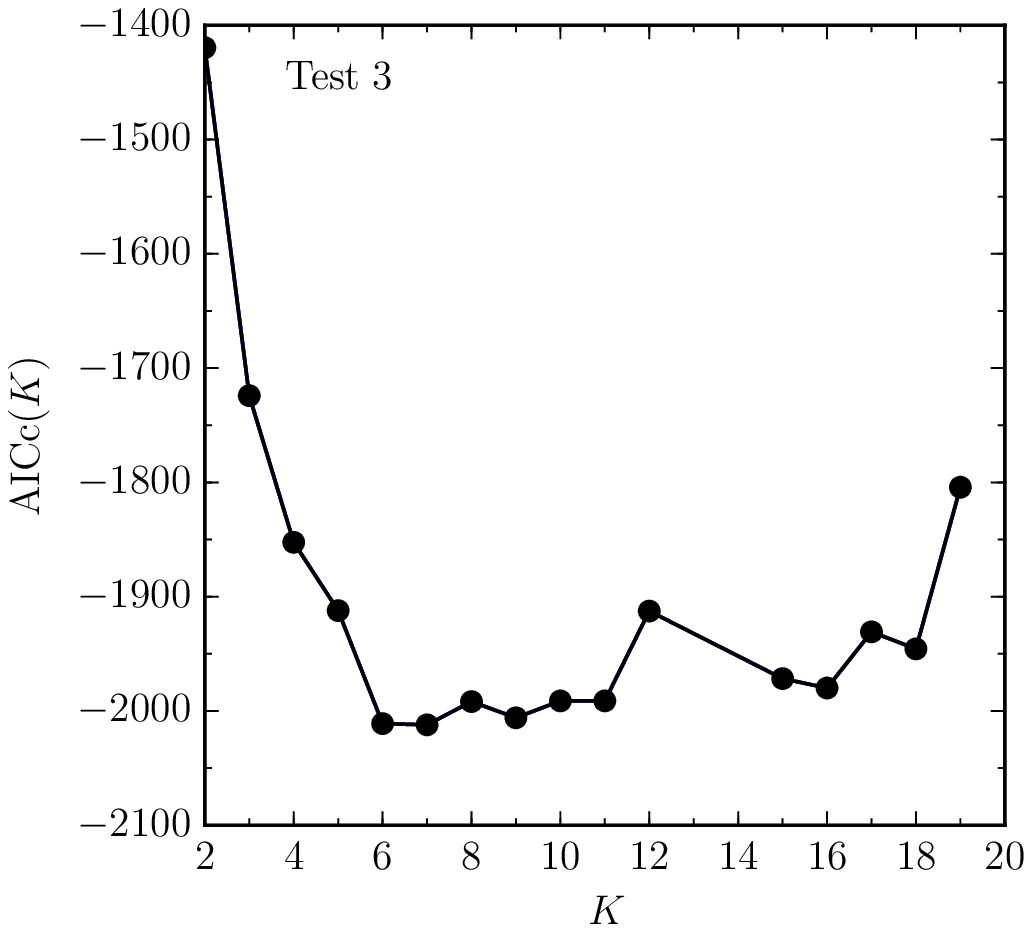}
\caption{The AICc as a function of the number ($K$) of Gaussian functions for the three simulation tests 
in Figure \ref{fig_sim}. The AICc reache a minimum at $K=13$, $10$, and $7$, respectively.}
\label{fig_aic}
\end{figure*}

\subsection{Markov-Chain Monte Carlo Implementation}
The parameters in our approach include two parameters ($\sigma$ and $\tau_d$) pertaining to
the DRW process in Equation~(\ref{eqn_cov_cc}) and $2K+1$ parameters ($f_k$s, $\tau_k$s, and $\omega$) pertainng
to the transfer function in Equation (\ref{eqn_tf}). To relax parameter degeneracy,
we introduce the following simplifications.
\begin{itemize}
 \item The central lags $\tau_k$ of Gaussian responses are assigned by a uniformly spaced grid of time lag 
 over a range of interests, whereas the amplitude $f_k$ is free to determine.
 Specifically, we set $\tau_k = \tau_1 + (k-1)\Delta \tau$ for $1\leqslant k \leqslant K$, 
 where $\Delta \tau$ is the grid space and $(\tau_1, \tau_K)$ is the range of
 the time lag. Given with the range of the time lag, the value of $\Delta \tau$ is determined by 
 the number ($K$) of Gaussian functions, i.e., $\Delta \tau = (\tau_K-\tau_1)/(K-1)$. 
 The degree of freedom is now reduced to $K+3$.

 \item The width of Gaussian responses $\omega$ is limited by two facts. If $\omega$ is too 
 small, unnecessary structures in transfer function will become admissible, and hence, the reconstructed
 light curves will appear to be unrealistically noisy; if $\omega$ is too
 large (e.g., $\omega>\Delta \tau$), the neighboring Gaussian responses are indeed indistinguishable. In practice, 
 we find it is adequate to adopt a prior limit of $\omega\sim (\Delta \tau/2, \Delta \tau)$.
\end{itemize}
The prior probability $P(\sigma, \tau_{\rm d}, \boldsymbol{\theta})$ in Equation (\ref{eqn_post}) 
is assigned by assuming that all the free parameters are independent of one another. The Gaussian width 
has a uniform prior, and the rest of the parameters have a logarithmic prior, which is a usual choice 
for ``scale'' parameters that involve a wide range of values
(e.g., \citealt{Sivia2006}, p. 108).

Lastly, we need to choose the number of Gaussian response functions.
Strictly speaking, one should employ the Bayesian model selection to 
choose the best parameter $K$, i.e., the one that maximizes the 
probability $P(K|\mathbi{y})$. However, it is very computationally expensive to calculate $P(K|\mathbi{y})$ since
one needs to marginalize all the other free parameters. Therefore, we resort to the relatively more economic 
``non-Bayesian'' approaches based on the maximum-likelihood estimate of the parameters (e.g., \citealt{Hooten2015}). 
We use the Akaike information criteria (AIC; \citealt{Akaike1973}), measures of the relative 
quality of statistical models.
The original form of the AIC is valid only asymptotically; \cite{Hurvich1989} proposed a correction to 
AIC for finite sample sizes
(abbreviated to AICc). The AIC is defined by
\begin{equation}
{\rm AIC} = 2m - 2\log P(\mathbi{y}|\hat\sigma,\hat\tau_{\rm d},\boldsymbol{\hat \theta}),
\end{equation}
where $m$ is the number of parameters in the model and the hat operator denotes the best estimate 
for the corresponding parameter. The AICc is defined by
\begin{equation}
{\rm AICc} = {\rm AIC} + \frac{2m(m+1)}{n-m-1},
\end{equation}
where $n=N+M$ is the total number of data points.  The best choice of $K$ is the one that minimizes the AICc.

We employ an MCMC method to determine the best estimates and uncertainties of 
free parameters. We use the Metropolis-Hastings algorithm to construct samples of free parameters
from the posterior probability distribution. The Markov chain is run in 150,000 steps in total.
The best estimates of free parameters 
are assigned the mean values of the samples, and the associated uncertainties are assigned
the 68.3\% confidence levels (``1$\sigma$'') that enclose the mean values.
The uncertainties of the final obtained transfer function in Equation~(\ref{eqn_tf})
are determined by the error propagation formulae, under the best inferred $K$.
Therefore, the uncertainty in the best inferred $K$ is not included.

\section{Simulation Tests}
In order to test our approach, we generate mock light curves of the continuum and broad emission lines with an
assumption of the transfer function, and then implement the above procedures to obtain the most probable
estimate for the transfer function and compare it with the input one. 
As outlined by \cite{Rybicki1992}, an unconstrained realization of mock light curves is simply 
a random series that has the covariance matrix $\mathbi{C=S+N}$ with specified parameters 
$\sigma$ and $\tau_{\rm d}$. We generate random numbers with the covariance matrix 
$\mathbi{C}$ using the Cholesky decomposition
(see \citealt{Zu2011}). The Cholesky decomposition of a matrix is $\mathbi{C=MM}^T$, where 
$\mathbi{M}$ is a lower triangular matrix with real and positive diagonal entries. 
If $\mathbi{r}$ is a series of Gaussian random numbers with a zero mean and unity dispersion, 
then $\mathbi{u=Mr}$ has the covariance matrix of $\mathbi{C}$, since $\langle\mathbi{uu}^T\rangle
=\mathbi{M}\langle\mathbi{rr}^T\rangle\mathbi{M}^T=\mathbi{MM}^T=\mathbi{C}$.
We first generate a mock continuum light curve and then convolve it with the adopted 
transfer function to produce a light curve of emission line.

\subsection{Validity of the Method}

We run three simulation tests, presuming the transfer functions to be composed of 
\begin{enumerate}
 \item two displaced top-hats,
 \item two moderately mixed Gaussians (so that the transfer function is double-peaked), and
 \item two severely mixed Gaussians, 
\end{enumerate}
respectively. The left panels of Figure~\ref{fig_sim} plot these three input transfer functions.
Specifically, for the first test, the two displaced top hats are (arbitrarily) configured to 
\begin{equation}
\Psi(\tau) = \frac{1}{15}\times\left\{
             \begin{array}{clc}
              1  & \rm~for & 5<\tau<10,\\
              2  & \rm~for & 15<\tau<20,\\
              0  & \rm~else,&
             \end{array}
             \right.
\end{equation}
where 1/15 is the normalization factor. 
For the other two tests, the two Gaussians are (arbitrarily) configured
to center at 10 and 18 days for the moderately mixed case
and 10 and 15 days for the severely mixed case, with the same standard deviation of 2.5 days. 
The mean time lags for these three transfer functions 
are 14.2, 14.0, and 12.5 days, respectively, roughly corresponding to a 5100~{\AA} luminosity 
of $\sim10^{43}~{\rm erg~s^{-1}}$ according 
to the relation between the BLR sizes and 5100~{\AA} luminosities of AGNs (\citealt{Bentz2013}).
For the DRW model of all the three tests, the parameter of the damping timescale is set to 
$\tau_{\rm d}=80$ days, a typical value for a luminosity of $10^{43}~{\rm erg~s^{-1}}$ 
(\citealt{Kelly2009, Zu2011, Li2013}); the parameter of the variation amplitude is set 
to $\sigma_{\rm d}=1.5$ (in arbitrary units) such that the resulting variation 
amplitude is typically $\sim$30-50\%. The time span of continuum monitoring is 150 
days, and the monitoring of the emission line starts 20 days latter. The cadence is typically 
one day apart. The signal-to-noise ratio (S/N) is set to $\sim$120.
It is worth stressing that the choice of these values is only for illustrative 
purposes and has no special meaning. The left panels of Figure~\ref{fig_sim} show the generated mock 
light curves of the continuum and emission lines for the three simulation tests. 

Throughout the calculations, the range of the time lag is set to (0-30) days. 
Since the best number of Gaussians is determined by the AICc, the final transfer 
function is insensitive to the adopted range of time lag,
provided it is broad enough to enclose the time lag of the emission line. 
The obtained AICc with the number of Gaussian functions $K$ for the three 
tests are illustrated in Figure~\ref{fig_aic}. 
The resulting optimal $K$ are 13, 10, and 7, respectively. This means that 
the optimal $K$ depends not on the sampling cadence but on the shape of the 
transfer function.
The right panels of Figure~\ref{fig_sim} show the best recovered transfer functions
with the optimal $K$.  
As can be seen, the structures in the transfer functions are successfully recovered,
and all the features in the light curves are also well reconstructed. 
However, we note that, in the first test, we fail to reproduce the top-hat shape 
because of the finite sampling and measurement errors of the light curves. 
Meanwhile, in the second test, although the recovered two peaks are marginally equal, 
the left peak seems slightly stronger than the right one.
These tests suggest that, like the performance of all other RM methods, that of the 
present approach depends on the data quality and the presence of strong variability 
features in the light curves; therefore, our approach may be unable to resolve very fine structures 
of the transfer function in some circumstances.

On the right panels of Figure~\ref{fig_sim}, 
we also superimpose the interpolation cross-correlation function (CCF) of the light curves 
using the procedure of \cite{Gaskell1987}. 
All the CCFs in the three tests are single-peaked, indicating that CCF analysis hardly 
reveals more than the characteristic time lags of BLRs. 
In particular, the CCF in the first test is wide and peaks at $\tau=17.5$ day.
This is an appropriate choice for the time lag if one only concentrates only on a single characteristic 
lag for the BLR, but clearly a single lag is not sufficient to describe the BLR in this case.
In Appendix C, we show the time lag distributions from the \texttt{JAVELIN} method
for the above three simulation tests. \texttt{JAVELIN} presumes a top-hat transfer function, 
and the time lag is assigned the mean lag of the top hat. We illustrate that 
while the present approach works fairly well in all the three tests, \texttt{JAVELIN}
cannot appropriately recover the characteristic lag for the first test with a multimodal 
transfer function.

\begin{figure}[t!]
\centering
\includegraphics[width=0.5\textwidth]{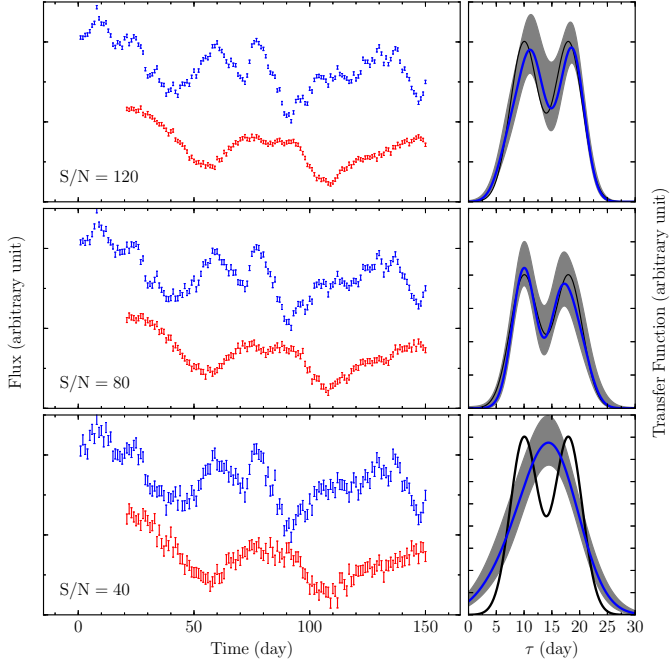}
\caption{Tests of our approach with S/N=120, 80, and 40. The input transfer function is 
presumed to be two moderately mixed Gaussians (see 3.1 for the details). 
For each test, the left panel shows the simulated light curves of the 5100{\AA} continuum (in blue) and 
H$\beta$ line (in red), and the right panel compares the best recovered transfer function (in blue) with the input one (in black)
transfer function. The shaded areas represents the estimated uncertainties for the 
recovered transfer function.}
\label{fig_sntest}
\end{figure}

\begin{figure}[t!]
\centering
\includegraphics[width=0.5\textwidth]{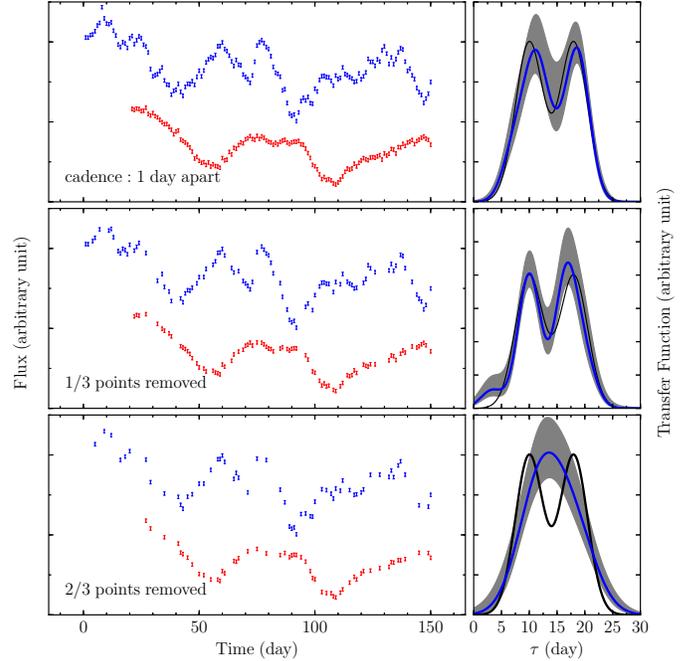}
\caption{Same as Figure \ref{fig_sntest} but for different cadences. On the top panel, the cadence is
1 day apart. On the middle and bottom panels, we randomly remove about 1/3, and 2/3,
respectively, of the points.}
\label{fig_cadencetest}
\end{figure}

\begin{figure}[t!]
\centering
\includegraphics[width=0.5\textwidth]{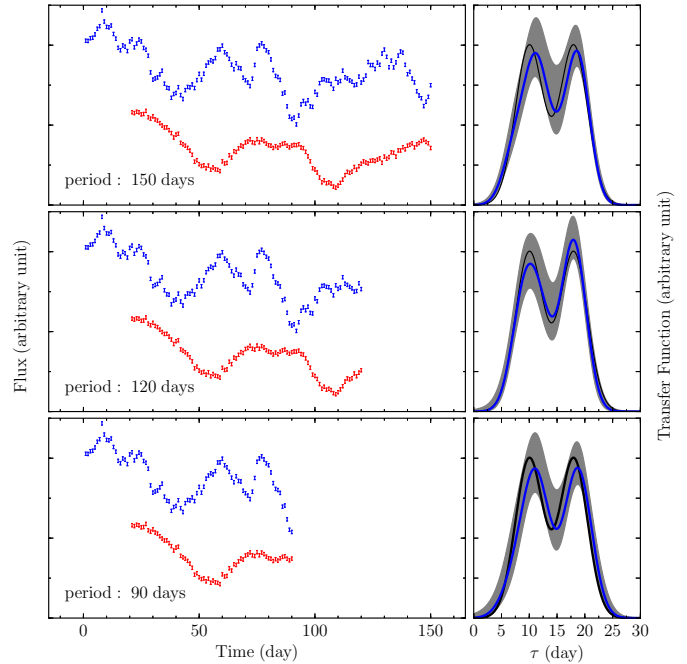}
\caption{Same as Figure \ref{fig_sntest} but for monitoring periods of 150, 120, and 90 days.}
\label{fig_spantest}
\end{figure}

\subsection{Dependence on Data Quality}
The above simulations show that our approach can generally well reproduce the input transfer 
function with optimal data quality. In this section, we test the dependence of 
the obtained transfer function on data quality with different S/N, sampling cadences,
and time spans of the mock light curves. We use for illustration only the double-peaked transfer function---namely, the one used for the second simulation test in the preceding section. 
We fix the seed for the random number generator so that all the mock light curves in this section
have the same variation pattern. This allows us to focus only on the desired factor that 
influences the obtained transfer function.

In Figure~\ref{fig_sntest}, we generate three sets of mock light curves with S/N=120, 80, and 40, respectively.
As expected, when S/N is low enough, the weak variability signals
in the light curves are overwhelmed by noises, and the information of small structures in the 
transfer function is missing. Consequently, for the case of S/N=40 (bottom panels of Figure~\ref{fig_sntest}), 
we obtain only a single-peaked transfer function, which has the same mean time lag 
as the input one.

To check how the sampling rate affects the results, we respectively randomly remove about 
one-third and two-thirds of the points of the light curves with S/N=120 in Figure~\ref{fig_sntest} 
and generate two new mock data sets with degraded sampling rates. We show the light 
curves and the corresponding best recovered transfer function in Figure~\ref{fig_cadencetest}.
Again, when the sampling cadence is poorer, more information on the fine structure 
of the transfer function is lost.

The default time span of the light curves is 150 days. In Figure~\ref{fig_spantest}, we 
reduce the time span to 120 and 90 days, by subtracting the last 30 and 60 days section,
of the light curves with S/N=120 in Figure~\ref{fig_sntest}, respectively.
The mean time lag of the input transfer function is 14 days, so that the overall time spans are still 
more than six times the mean lag (\citealt{Welsh1999}).
There is no surprise that the transfer function is well reproduced in all the three cases
in Figure~\ref{fig_spantest}.

In summary, provided the time span is more than several times the mean time lag of the 
emission line, the S/N and sampling cadence are more critical to recovering the 
transfer function. We stress again that the overall performance of the 
present approach depends on the presence of strong variation features in the light curves
and the specific structures of the transfer function one desires to resolve. 
We therefore cannot specify more details about the data requirements for the present approach
to faithfully recover the transfer function.

\begin{figure*}[th!]
\centering
\includegraphics[width=0.75\textwidth]{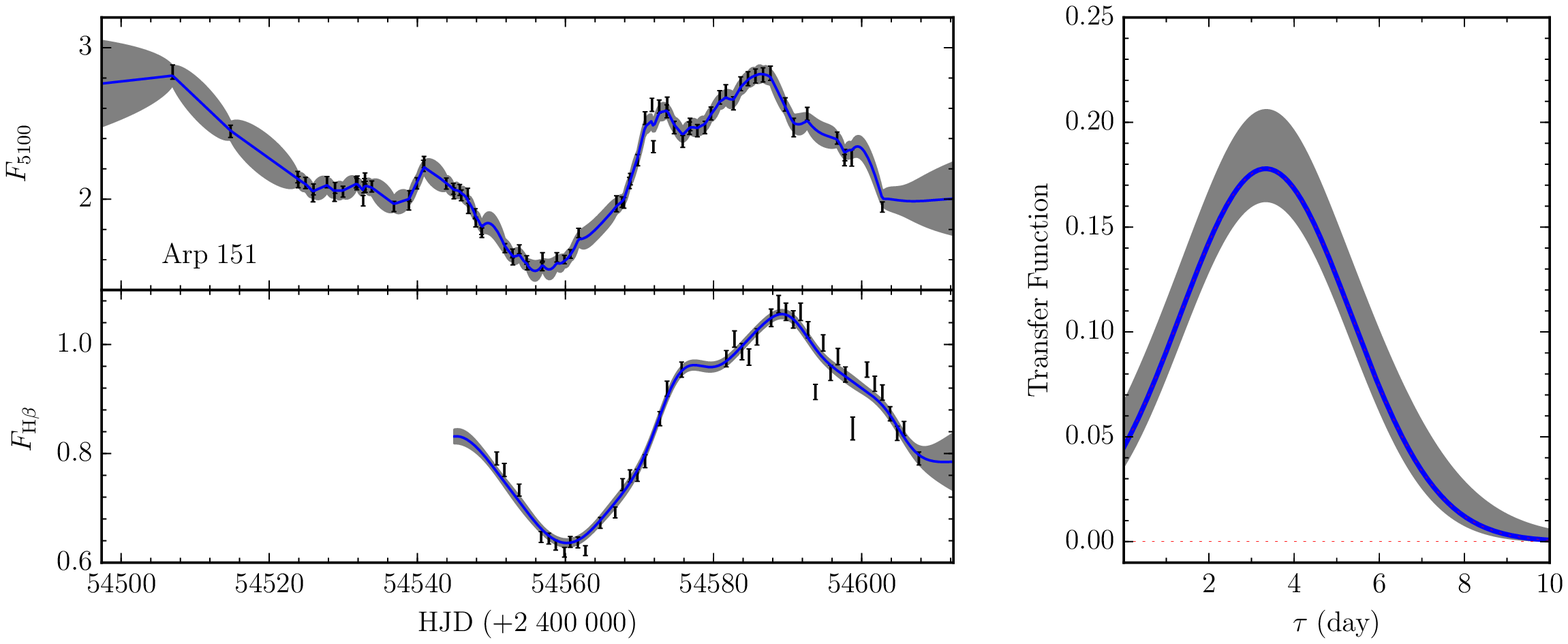}
\includegraphics[width=0.75\textwidth]{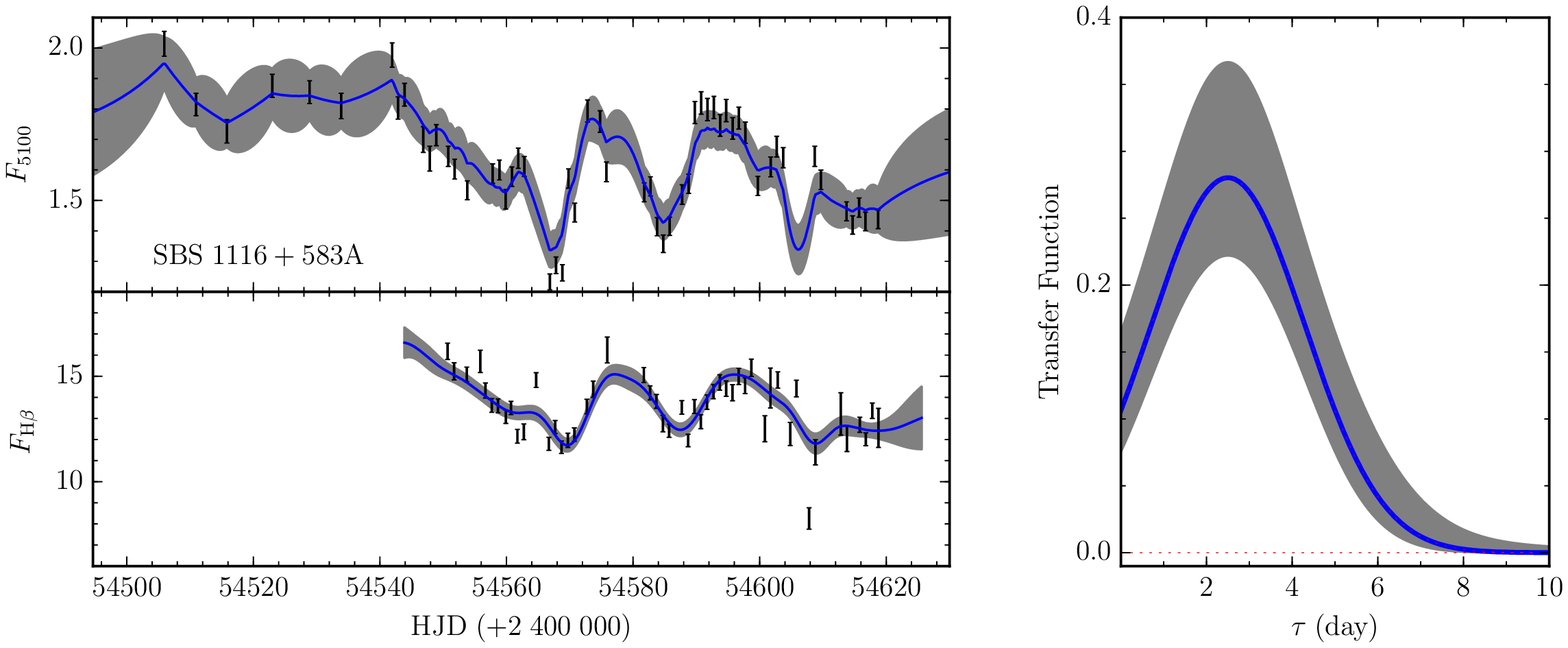}
\includegraphics[width=0.75\textwidth]{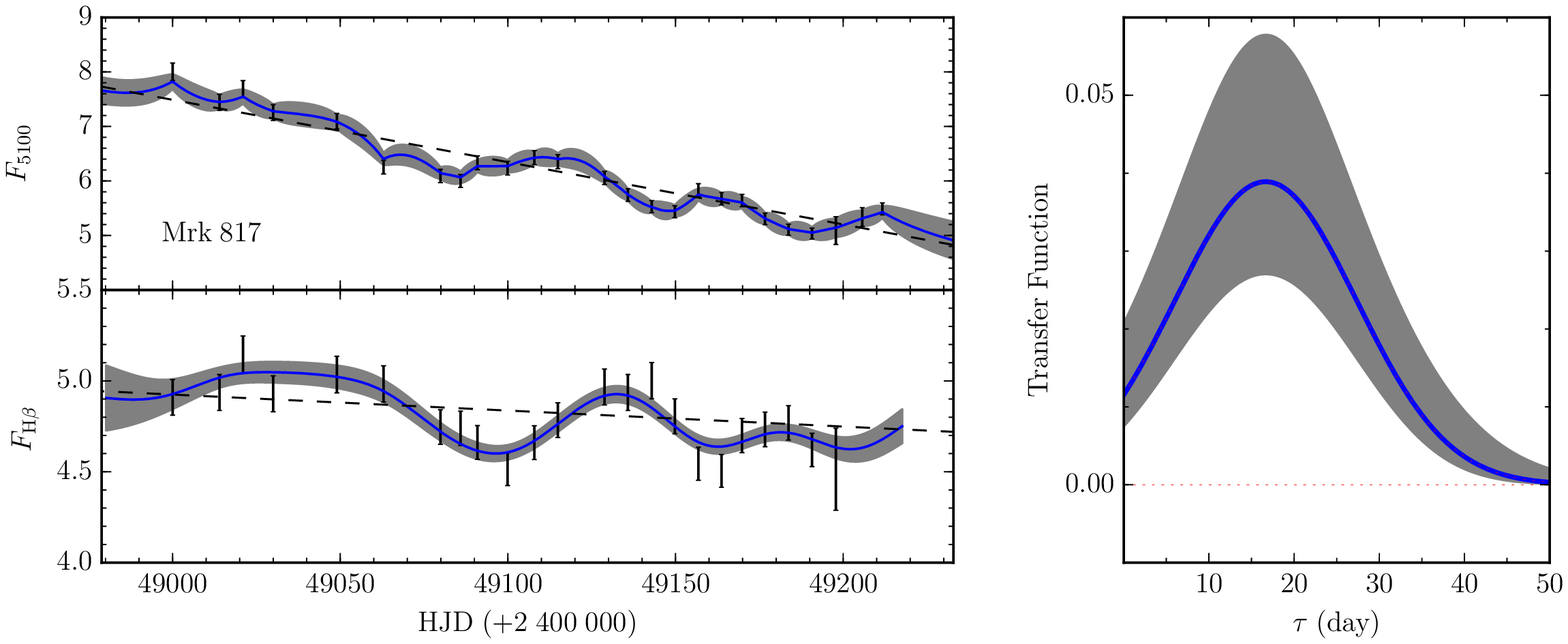}
\caption{The best recovered transfer functions for Arp 151 (top), SBS~1116+583A (middle), 
and Mrk 817 (bottom).
For each object, the two left panels show the light curves of continuum and H$\beta$ line (points with error
bars), together with the best reconstructions (solid lines with shaded areas). The right panel shows the 
best recovered transfer function, with the shaded area representing the uncertainties. Red dotted 
lines represent the Gaussian functions that constitute the transfer function. 
For Mrk~817, a detrending of the long-term trend is included. The dashed lines in the two left panels show 
the best fit for the detrending.}
\label{fig_example}
\end{figure*}

\section{Application}
We select three illustrative RM sources to show the feasibility of our approach: 
Arp 151, SBS~1116+583A, and Mrk~817.
The first two objects were monitored by the Lick AGN Monitoring Project (LAMP
\footnote{The spectroscopic data of LAMP project are publicly available from 
\url{http://www.physics.uci.edu/~barth/lamp.html}.}%
; \citealt{Bentz2009, Walsh2009}), and 
the last object was monitored by \cite{Peterson1998}. The RM data of Arp 151 were previously
studied in great detail using the maximum entropy method (\citealt{Bentz2010}), BLR dynamical 
modeling method (\citealt{Brewer2011, Pancoast2014b}), and RLI method 
(\citealt{Skielboe2015}). Arp 151 therefore serves as a good example to further verify the validity of our approach. 
The transfer function of SBS 1116+583A derived by the RLI method 
of \citeauthor{Skielboe2015} interestingly showed a bimodal distribution, meriting an analysis 
using our approach.  The object Mrk~817 shows long-term secular trends in its light curves
(\citealt{Li2013}) and is therefore adopted to illustrate the capability of our approach 
for coping with detrending. 

To compare our results with those of CCF analysis, we define the mean time delay 
from the transfer function as (\citealt{Rybicki1994})
\begin{equation}
\tau_{\rm m}=\frac{\int_{-\infty}^\infty \tau \Psi(\tau)d\tau}{\int_{-\infty}^\infty \Psi(\tau)d\tau}
=\frac{\sum_k f_k\tau_k}{\sum_k f_k}.
\label{eqn_lag}
\end{equation}
The uncertainty of $\tau_{\rm m}$ is calculated by the error propagation formulae.

\begin{figure}[thp!]
\centering
\includegraphics[width=0.35\textwidth]{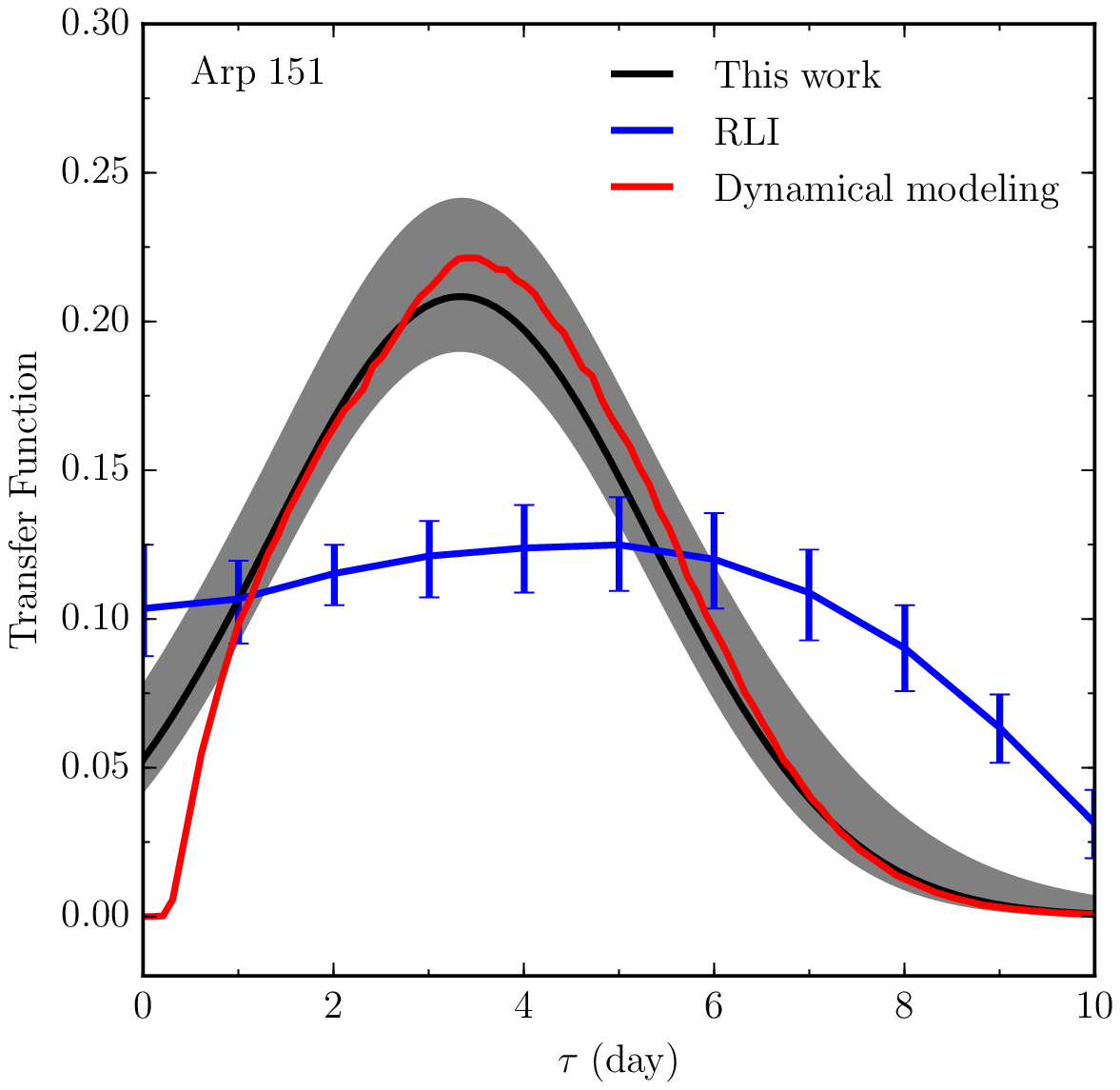}
\includegraphics[width=0.35\textwidth]{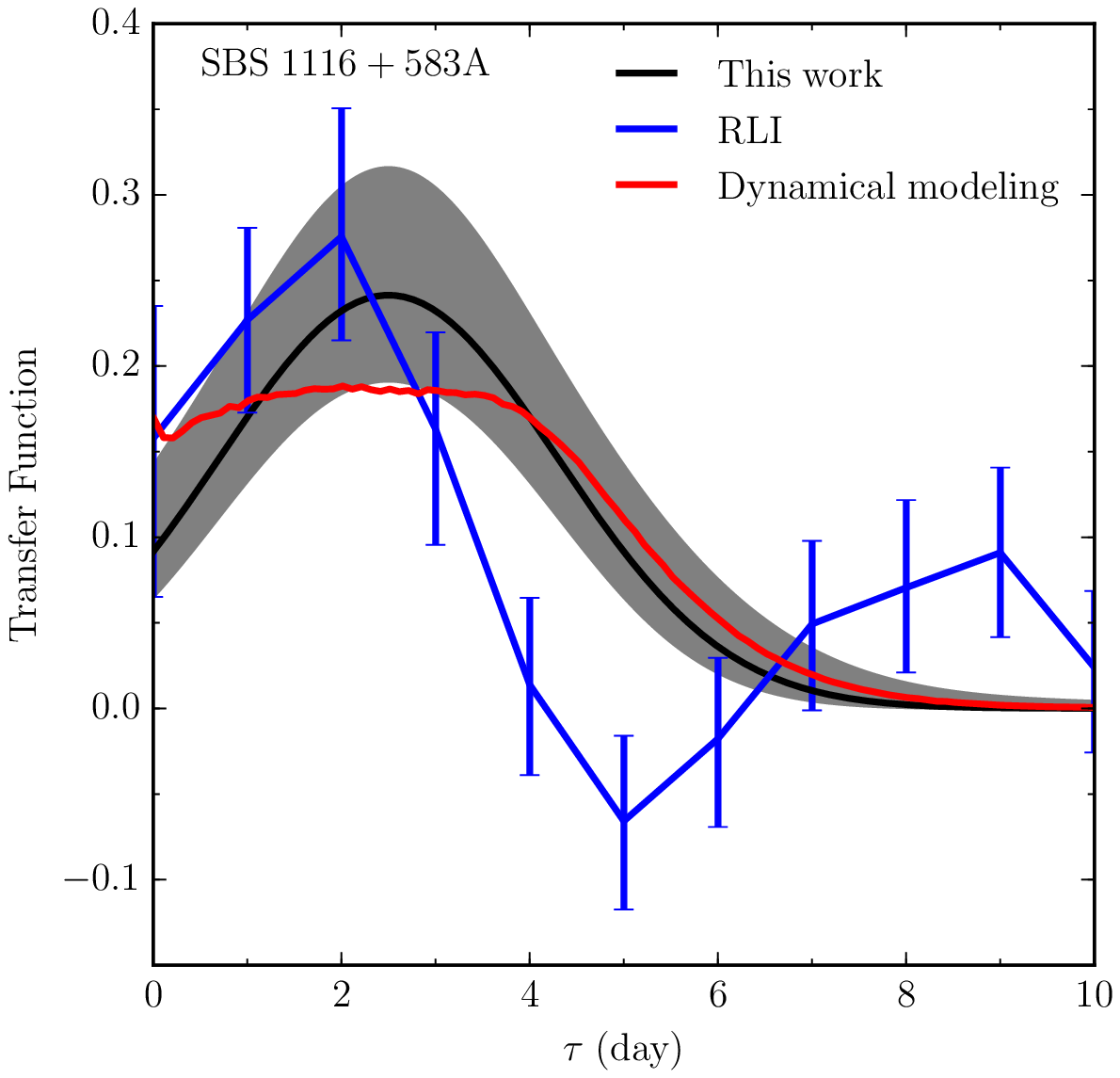}
\includegraphics[width=0.35\textwidth]{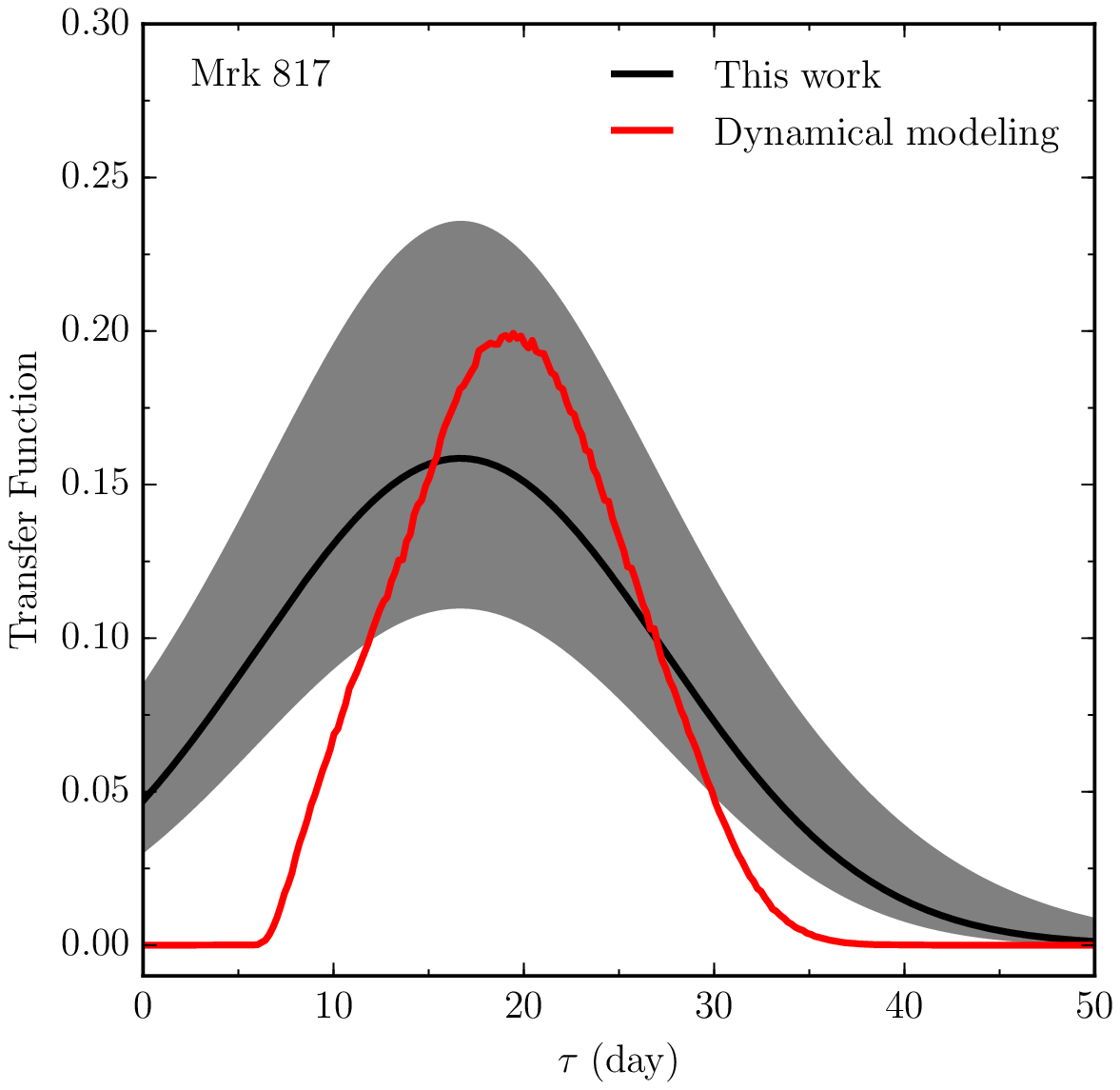}
\caption{Comparisons of the transfer functions derived from different methods for Arp~151 (top), 
SBS~1116+583A (middle), and Mrk~817 (bottom). The black line with shaded areas is the transfer function
from this work, the blue line with error bars is from the regularized linear inversion (RLI) method
of \cite{Skielboe2015}, and the red line is from the BLR dynamical modeling method of \cite{Li2013}.
Note that there is no RLI solution for Mrk~817. All the transfer functions are normalized 
on a common scale.}
\label{fig_tf}
\end{figure}

\subsection{Arp~151}
For the sake of comparison with the previous results, we use the $B-$band photometric light 
curve of Arp~151 as a surrogate of the continuum as in \cite{Pancoast2014b} and \cite{Skielboe2015}.  
The magnitudes are first converted into fluxes. Since the absolute unit of light 
curves is no longer important for RM analysis, we adopt an arbitrary zero-magnitude flux 
for the conversion. We set the range of the time lag for solving the transfer function to be
(0-10) days. The best number of Gaussian functions is $K=4$ according the AICc.

The top panels of Figure~\ref{fig_example} present the reconstruction 
for the light curves of the continuum and H$\beta$ line and the best recovered transfer
function for Arp~151. The mean time lag from Equation (\ref{eqn_lag}) 
is $\tau_{\rm m} = 3.33\pm0.26$ days, consistent with the previous results from CCF analysis,
$\tau_{\rm CCF}=4.08\pm0.60$ days (\citealt{Bentz2009}), and 
the \texttt{JAVELIN} method of \cite{Zu2011}, $\tau_{\rm JAV}=3.68\pm0.36$ days (\citealt{Grier2013a}).
The transfer function is mainly contributed by the second Gaussian component centered at 
$\tau=3.33$ days. 

On the top panel of Figure~\ref{fig_tf}, we compare the transfer function 
derived in this work with those derived from the RLI method
(\citealt{Skielboe2015}) and from the BLR dynamical method (\citealt{Li2013}).
Our result well matches that from the dynamical modeling method but differs from 
the RLI result, which has a plateau from 0 to about 7 days. 
The shape of our obtained transfer function also coincides with that from 
the maximum entropy method (see Figure~1 in \citealt{Bentz2010}).

\subsection{SBS~1116+583A}
As we have done for for Arp~151, we use the $B-$band photometric light curve for this object.
The range of the time lag in the calculations is set to (0-10) days, and the best number of 
Gaussian functions is $K=5$ according to the AICc. The results are shown
on the middle panels of Figure~\ref{fig_example}. The obtained mean time lag 
is $\tau_{\rm m}=2.50\pm0.70$ days, which is again well consistent with 
the results based on CCF analysis, $\tau_{\rm CCF}=2.38\pm0.58$ days (\citealt{Bentz2009}), and the \texttt{JAVELIN} 
method, $\tau_{\rm JAV}=2.46\pm0.93$ days (\citealt{Grier2013a}).

The middle panel of Figure~\ref{fig_tf} compares the derived transfer functions
from various methods. The RLI method yields a bimodal transfer function
with a response component of around 2 days and an additional component  of around 8 days
(\citealt{Skielboe2015}). However, both the transfer functions obtained in
this work and those from the dynamical modeling do not show the second component. An inspection of 
the reconstructed H$\beta$ light curves (see Figure 10 in \citealt{Skielboe2015}) 
indicates that our results better match the observations.  

\subsection{Mrk~817}
Mrk~817 was monitored for as long as five years by \cite{Peterson1998}, but only
during the intervening three years were H$\beta$ time lags detected.  We 
use the first-year section  of the light curves with the H$\beta$ time 
lags detected (HJD 24,449,000$-$24,449,211). The CCF analysis by \cite{Peterson1998}
yielded a time lag of $\tau_{\rm CCF}=20.1\pm4.4$ days.
The range of the time lag in our calculations is set to (0-50) days and 
the best number of Gaussian functions is found to be $K=4$. 
It is apparent from a simple visual inspection that the light curves of 
the continuum and H$\beta$ line show different long-term trends, with that of the continuum fluxes
decreasing more steeply. We therefore include the detrending procedure as described in Section~2.2,
in which both the trends of both the continuum and the H$\beta$ line are modeled by a linear polynomial.
The results are plotted on the bottom panels of Figure~\ref{fig_example}.
The transfer function peaks at around 17 days and bears large uncertainties due to poor data sampling.
The resulting mean time lag also has a fairly large uncertainty $\tau_{\rm m}=16.7\pm8.3$ days.
Nevertheless, all the variation features in the light curves are well reproduced.

This object has never been analyzed by the RLI method or the maximum entropy method.
The bottom panel of Figure~\ref{fig_tf} only shows a comparison of our result with only that of 
the dynamical modeling method. The two results are generally
consistent, although our result peaks at a slightly shorter time lag.

In our application to the above three RM objects, the transfer functions can be 
effectively modeled by a single Gaussian. As illustrated in Section~3.2, the time resolution 
of the obtained transfer function
is determined by the data sampling and measurement noises of the observations.
In a statistical sense, based on the AICc criteria, the present solutions are the best ones that
balance the trade-off between the goodness of fit  and the complexity of our approach.
Refining the sampling and signal-to-noise ratio can certainly provide more constraints on 
the fine structures in transfer functions (see Figure~\ref{fig_sch}) if they are not
intrinsically a simple single Gaussian.

\section{Discussion and Conclusions}

By extending the previous work of \cite{Rybicki1992} and \cite{Zu2011}, we 
describe a non-parametric approach to determine the transfer function in RM.  
The basic principle that the transfer function can be expressed as a sum of a family of relatively
displaced Gaussian functions enables us to largely obviate the need for the presumption of 
a specified transfer function or BLR model as in previous studies. Thus, this to some extent makes our 
approach model independent. In addition, following the framework of \cite{Zu2011}, 
the inclusion of the statistical modeling of continuum 
variations as a DRW process allows us to naturally take into account the measurement errors and make 
use of the information in observation data sets as much as possible. 
The application of our approach to RM data sets illustrates its fidelity of our approach and its 
capability to deal with long-term, secular variations in light curves.
In particular, our approach is apt for solving the multimodal transfer functions
associated with multiple-component structures of BLRs (e.g., \citealt{Hu2012}).
In addiation to BLRs, our approach can also be applied to RM observations of dusty tori in AGNs 
(e.g., \citealt{Suganuma2006,Koshida2014}).

Since we relax the presumption of a specified transfer function but let its shape 
be determined by the data, our approach can be regarded as a useful complementary to the BLR dynamical modeling (\citealt{Pancoast2011, Pancoast2014a, Li2013}).
Indeed, interpretation of the obtained transfer functions  has to invoke physical BLR
models. A combination of these two methods may help to remove the concern over the systematic errors 
in the dynamical modeling. On the other hand, in our approach (compared with the RLI method and 
the maximum entropy method), we incorporate the statistical description for the light 
curves of both the continuum and the emission line; therefore, we take full advantage of the information
in the data and can estimate uncertainties for the obtained transfer function in a self-consistent
manner. 

We end with discussions on some potential improvements to be made to our approach. 

\begin{itemize}
 \item Throughout our calculations, we have by default assumed a linear response of the emission line to 
 the continuum so that the covariance functions of the light curves can be analytically expressed
 by the error function. To include nonlinear response (\citealt{Li2013}), we can introduce 
 a new parameter $\gamma$ for the emission line fluxes as
 \begin{equation}
 f_l(t) = f_{l, \rm new}^{1+\gamma}(t),
 \end{equation}
 so that the new flux series $f_{l, \rm new}(t)$ now linearly responds to the continuum.
 
 \item Since our approach is based on the statistical modeling of the temporal variations, 
 it is not straightforward to directly include the velocity information. 
 We can apply our approach to each velocity bin individually (e.g., \citealt{Skielboe2015}).
 In this case, any correlation information between velocity bins is neglected; thus, the 
 obtained velocity-resolved transfer function may be noisy along the velocity direction.
 It merits a future development of a feasible scheme to self-consistently take into account
 the velocity information.
 
 \item We use the DRW process to describe the continuum variations, a process found to be sufficiently
 adequate for most of current AGN variability data sets (e.g., 
 \citealt{Kelly2009, MacLeod2010, Li2013, Zu2013}). However, there is evidence  
 that the DRW process is no longer appropriate on a short time scale of days (\citealt{Mushotzky2011,Kasliwal2015, 
 Kozlowski2016}). This impedes the application of our approach in recovering very fine 
 structures (on a timescale of days) in the transfer function.
 Recently, \cite{Kelly2014} 
 developed a flexible method that statistically describes the variability by 
 a general continuous-time autoregressive moving average (CARMA) process. The DRW process 
 is a special case of the CARMA process. In principle, our approach can also 
 naturally include the CARMA process, but this will make out approach highly computationally 
 expensive. We therefore defer this to a future improvement with the parallelization of
 our approach on supercomputer clusters.

\end{itemize}

\acknowledgements
We thank the two referees (astronomer and statistician) for their insightful suggestions, which improved the manuscript. We are grateful to the staff of the Lijiang station of the Yunnan Observatories for their assistance with the RM data sets.
YRL thanks Ying Zu for useful discussions and Andreas Skielboe for providing the data of the H$\beta$ light 
curves and their derived transfer function for SBS~1116+583A.
This research is supported in part by the Strategic Priority Research Program 
- The Emergence of Cosmological Structures of the Chinese 
Academy of Sciences, Grant No. XDB09000000, and by the NSFC (Gran No.  NSFC-11173023, -11133006, 
-11233003, -11303026, -11573026, and -U1431228). This work is also supported in part by grant 2016YFA0400700 from the Ministry of Science and Technology of China. This work has made use of data from the Lick AGN Monitoring Project public data release.

We developed for the present approach a software package \texttt{MICA} (Multiple and Inhomogeneous 
Component Analysis), which is available upon request.

\begin{figure*}[th!]
\centering
\includegraphics[width=0.3\textwidth]{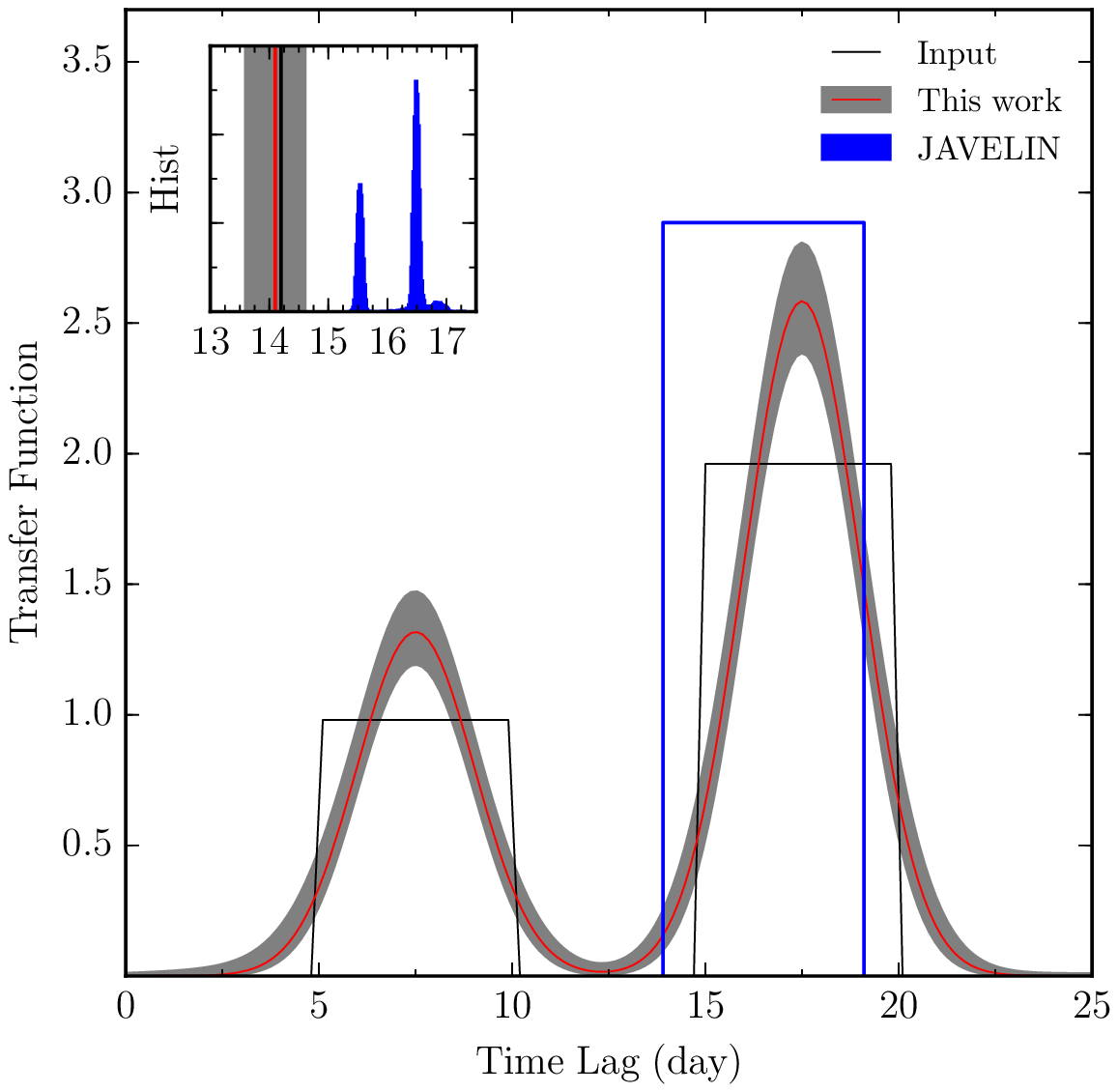}
\includegraphics[width=0.3\textwidth]{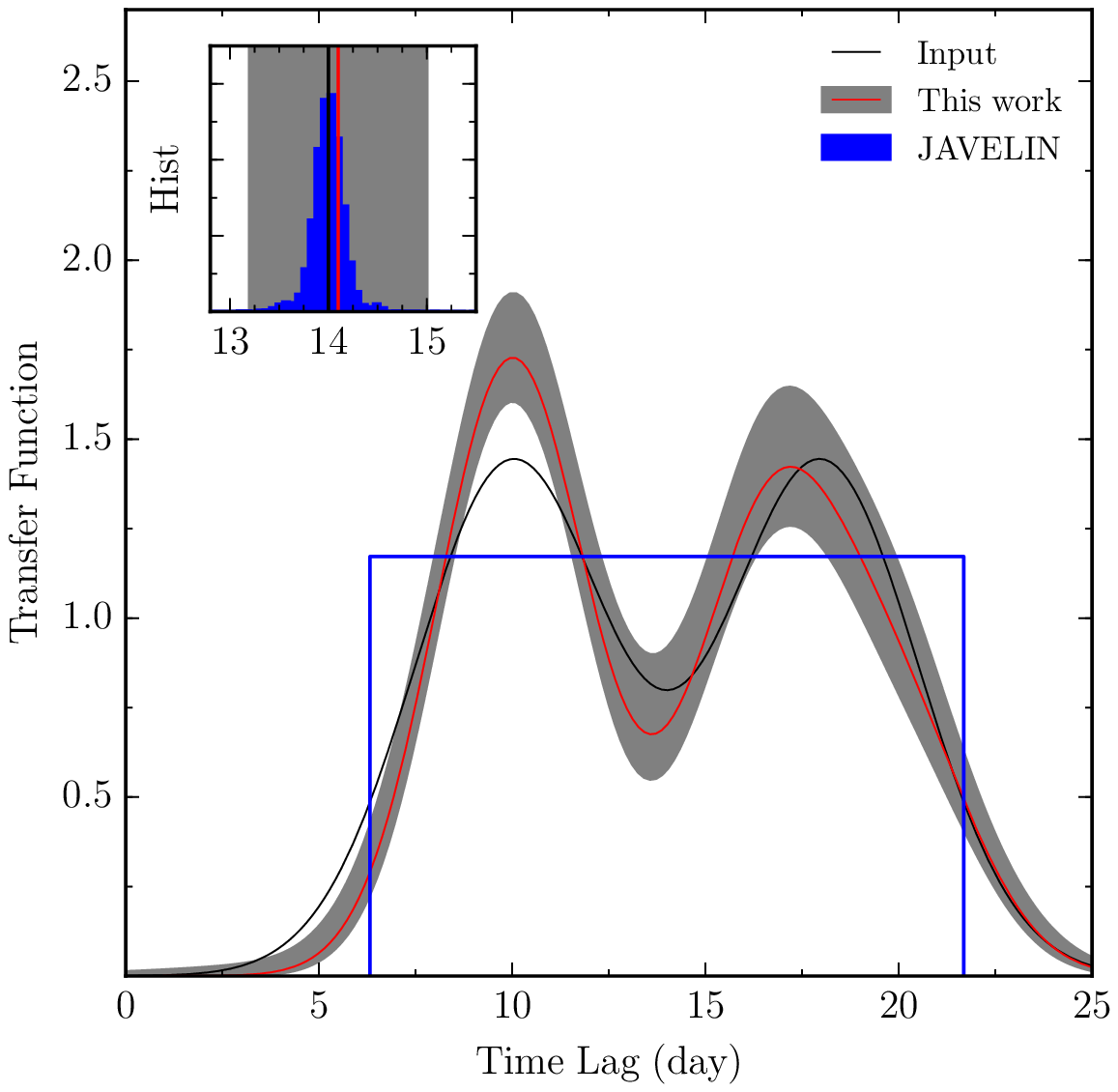}
\includegraphics[width=0.292\textwidth]{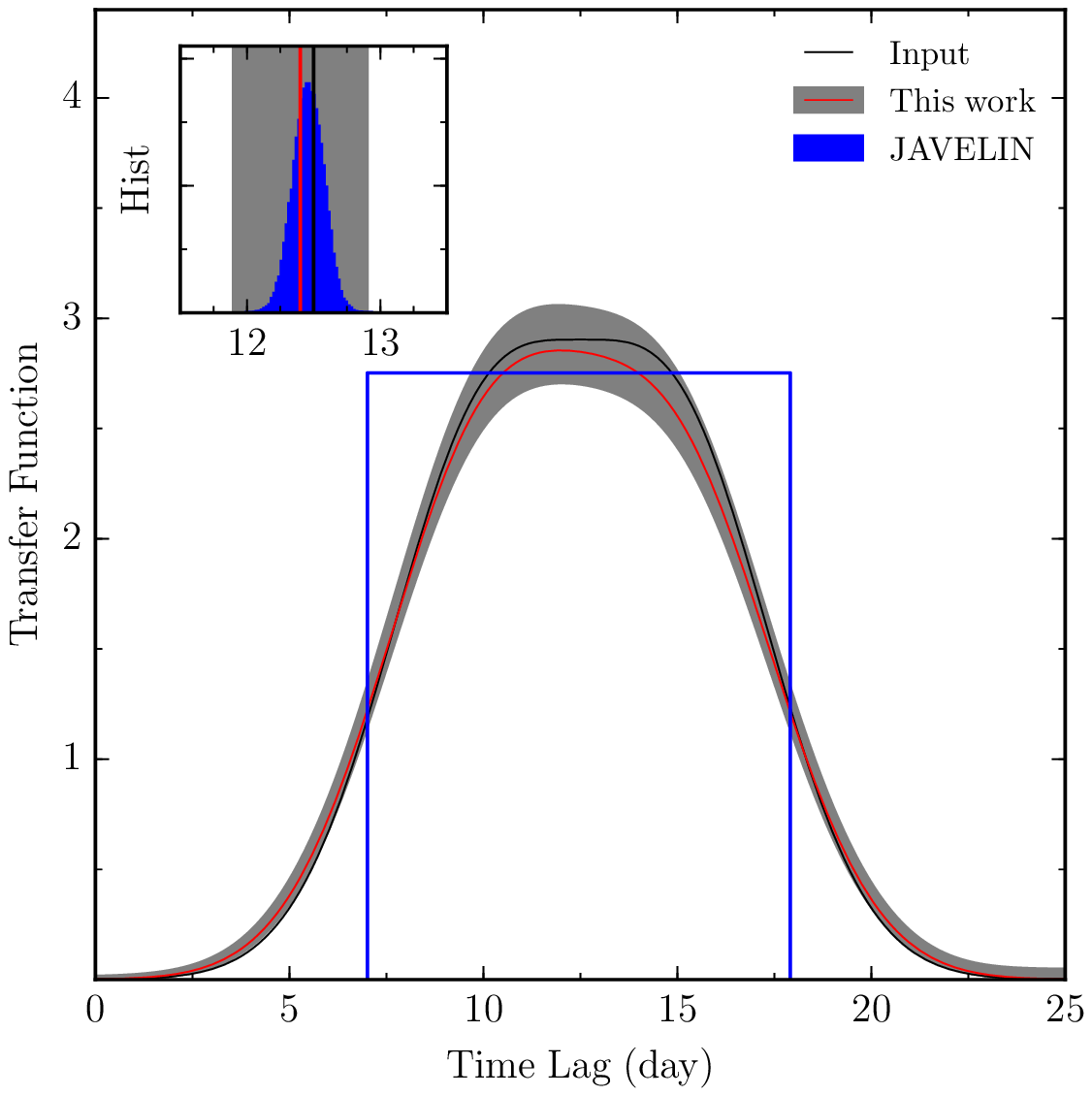}
\caption{ Comparison of the best recovered transfer functions from our approach (red lines with shaded areas)
with \texttt{JAVELIN} (blue lines) for the simulated light curves of the three tests in Figure~\ref{fig_sim}. 
Black lines show the input transfer functions. The transfer functions are adjusted to a common scale. 
The inset panels highlight the histograms
of the mean time lags of the top hat from \texttt{JAVELIN}, along with the mean time lags of the transfer 
functions from the input and our approach, indicated by black vertical lines and red vertical lines with
shaded areas, respectively.}
\label{fig_javelin}
\end{figure*}

\section*{Appendix A\\ Covariance functions}
In this appendix, we derive analytical expressions for $S_{lc}$ and $S_{ll}$ in 
Equations~(\ref{eqn_cov_lc}) and (\ref{eqn_cov_ll}).
In deriving $S_{lc}$, for the sake of clarity, we denote 
\begin{equation}
S_{lc}(\Delta t) = \sum_k S_{lc}^k(\Delta t),
\end{equation}
where $S_{lc}^k$ represents the contribution to the covariance from the $k$th Gaussian component.
It is easy to show
\begin{eqnarray}
S_{lc}^k(\Delta t)
&=& \sigma^2f_k\left[\int_{-\infty}^{\Delta t} + \int_{\Delta t}^\infty\right]
e^{-(\tau-\tau_k)^2/2\omega^2-|\tau-\Delta t|/\tau_d} d\tau\nonumber\\
&=&\sqrt{\frac{\pi}{2}}\sigma^2 \omega f_k e^{\omega^2/2\tau_{\rm d}^2} 
\left\{e^{-\Delta T_1/\tau_{\rm d}}
{\rm erfc}\left[-\frac{1}{\sqrt{2}}\left(\frac{\Delta T_1}{\omega}-\frac{\omega}{\tau_{\rm d}}\right)\right]\right.\nonumber\\
&&\left.+e^{\Delta T_1/\tau_{\rm d}}{\rm erfc}\left[\frac{1}{\sqrt{2}}
\left(\frac{\Delta T_1}{\omega}+\frac{\omega}{\tau_{\rm d}}\right)\right]
\right\},
\end{eqnarray}
where $\Delta T_1=\Delta t-\tau_k$ and ${\rm erfc}(x)$ is the complementary error function.

Similarly, we denote 
\begin{equation}
S_{ll}(\Delta t) = \sum_k\sum_m S_{ll}^{km}(\Delta t).
\end{equation}
The covariance component $S_{ll}^{km}$ is
\begin{eqnarray}
S_{ll}^{km}(\Delta t)&=&\sigma^2f_kf_m
\int_{-\infty}^\infty\int_{-\infty}^\infty e^{-(\tau-\tau_k)^2/2\omega^2} e^{-(\tau'-\tau_m)^2/2\omega^2}\nonumber\\  
&&\times e^{-|\Delta t - (\tau-\tau')|/\tau_{\rm d}} d\tau'd\tau.
\end{eqnarray}
Introducing new variables $u=(\tau + \tau')$ and $v=(\tau-\tau')$, we have
\begin{eqnarray}
S_{ll}^{km}(\Delta t)& =& \frac{1}{2}f_mf_k\sigma^2\int_{-\infty}^{\infty}du \int_{-\infty}^{\infty}dv e^{-(u+v-2\tau_k)^2/8\omega^2}\nonumber\\
 &&\times  e^{-(u-v-2\tau_m)^2/8\omega^2}
e^{-|\Delta t - v|/\tau_{\rm d}}.
\end{eqnarray}
Note that 
\begin{eqnarray}
&&(u+v-2\tau_k)^2+(u-v-2\tau_m)^2 \nonumber \\
&&=2\left[u-(\tau_k+\tau_m)\right]^2 + 2\left[v-(\tau_k-\tau_m)\right]^2.
\end{eqnarray}
The above integral can be rewritten as 
\begin{eqnarray}
S_{ll}^{km}(\Delta t) &=& \frac{1}{2}f_mf_k\sigma^2
\int_{-\infty}^\infty du e^{-[u-(\tau_k+\tau_m)]^2/4\omega^2}\nonumber\\
&&\times\int_{-\infty}^{\infty}dv e^{-[v-(\tau_k-\tau_m)]^2/4\omega^2 - |\Delta t - v|/\tau_{\rm d}}.
\end{eqnarray}
The first integration term is
\begin{equation}
\int_{-\infty}^\infty  e^{-[u-(\tau_k+\tau_m)]^2/4\omega^2} du = 2\sqrt{\pi}\omega,
\end{equation}
and the second integration term is
\begin{eqnarray}
\left[\int_{-\infty}^{\Delta t}\right.&+&\left.\int^{\infty}_{\Delta t}\right]
e^{-[v-(\tau_k-\tau_m)]^2/4\omega^2 - |\Delta t - v|/\tau_{\rm d}} dv \nonumber \\
&=&\sqrt{\pi}\omega e^{\omega^2/\tau_{\rm d}^2} 
\left\{ 
  e^{-\Delta T_2/\tau_{\rm d}}{\rm erfc}\left[-\frac{\Delta T_2}{2\omega}+\frac{\omega}{\tau_{\rm d}}\right]\right.\nonumber\\
&&\left.+ e^{\Delta T_2/\tau_{\rm d}}{\rm erfc}\left[\frac{\Delta T_2}{2\omega}+\frac{\omega}{\tau_{\rm d}}\right]
\right\},
\end{eqnarray}
where $\Delta T_2 = \Delta t - (\tau_k-\tau_m)$.
As a result, we arrive at
\begin{eqnarray}
S_{ll}^{km}(\Delta t) &=& \pi\omega^2f_mf_k\sigma^2 e^{\omega^2/\tau_{\rm d}^2}
\left\{ 
  e^{-\Delta T_2/\tau_{\rm d}}{\rm erfc}\left[-\frac{\Delta T_2}{2\omega}+\frac{\omega}{\tau_{\rm d}}\right]\right.\nonumber\\
&&\left.+ e^{\Delta T_2/\tau_{\rm d}}{\rm erfc}\left[\frac{\Delta T_2}{2\omega}+\frac{\omega}{\tau_{\rm d}}\right]
\right\}.
\end{eqnarray}

\section*{Appendix B\\ Error function}
The error function is defined as
\begin{equation}
{\rm erf}(x) = \frac{2}{\sqrt{\pi}}\int^x_0 e^{-t^2}dt.
\end{equation}
The complementary error function is defined as
\begin{equation}
{\rm erfc}(x) = 1-{\rm erf}(x) = \frac{2}{\sqrt{\pi}}\int_x^\infty e^{-t^2}dt.
\end{equation}
It is trivial to show that the integral 
\begin{equation}
\int^x_{-\infty} e^{-t^2}dt = \frac{\sqrt{\pi}}{2} {\rm erfc}(-x) ,
\end{equation}
and the intergal
\begin{eqnarray}
\int e^{-(at^2+bt+c)} dt &= &-\frac{1}{2}\sqrt{\frac{\pi}{a}}e^{(b^2-4ac)/4a} \nonumber\\
&&\times{\rm erfc}\left(\sqrt{a}x + \frac{b}{2\sqrt{a}}\right) + {\rm const.}
\end{eqnarray}

\section*{Appendix C\\ Comparison with \texttt{JAVELIN}}
The present approach extends the work of \cite{Zu2011} by 
adopting a more flexible transfer function rather than a specified top-hat transfer function.
Therefore, we can obtain more than the characteristic time lag and recover complicated structures 
in the transfer function. In Figure~\ref{fig_javelin}, we run the public package \texttt{JAVELIN}, developed 
by \cite{Zu2011}, on the simulated light curves of the three tests in Section~3.1 and plot the obtained
top-hat transfer functions and the distributions of the mean time lag of the top-hat. 
The recovered transfer functions 
from our approach and the input transfer functions are also superimposed for comparison.
\texttt{JAVELIN} also employs an MCMC method (but with a different algorithm) to determine the 
best values of the free parameters. Again, we run the Markov chain with 150,000 steps. 
The time-lag distribution for the first test is bimodal, 
with the two peaks at 15.53$\pm$0.04 and 16.5$\pm$0.1 days, respectively, and the overall mean lag of 
16.2$\pm$0.5 days.
Here, the uncertainties are assigned the standard deviations of the distributions.
The two peaks correspond to neither
the mean lags of the input two top hats (7.5 and 17.5 days) nor the overall mean lag (14.2 days).
The \texttt{JAVELIN} time lags for the second and third tests are 14.0$\pm$0.2 
and 12.5$\pm$0.1 days, respectively, 
in agreement with the mean lags of the input transfer functions (14.0 and 12.5 days). However,
for the second test, the time lag does not show the expected double-peak distribution.
By contrast, the present approach works fairly well in all these three cases (see Figure~\ref{fig_sim}),
in terms of recovering both the shape of the transfer functions and the characteristic time lags.
The present approach yields a mean time lag of 14.1$\pm$0.5, 14.1$\pm$0.9, and 12.4$\pm$0.5 days for the 
three tests, respectively,
generally consistent with the input values. Here, the uncertainties are calculated by applying 
the error propagation formulae to Equation (\ref{eqn_lag}).
We attribute such discrepancies of \texttt{JAVELIN} to the too simplified top-hat transfer function adopted
in the approach, which may encounter difficulties when the transfer function is multimodal and 
asymmetric.

\end{document}